%% file: ms.tex
\documentclass[a4paper,fleqn,usenatbib]{mnras}

\usepackage{newtxtext,newtxmath}

\usepackage[T1]{fontenc}
\usepackage{ae,aecompl}

\usepackage{graphicx} 
\usepackage{amsmath}  
\usepackage{amssymb}  

\usepackage{amsfonts}
\usepackage{xspace}
\usepackage{gensymb}
\usepackage{subfigure}
\usepackage[usenames, dvipsnames]{color}
\usepackage{times}
\newcommand{\co}{CO\xspace}

\newcommand{\cotwoone}{\mbox{CO(2-1)}\xspace}

\newcommand{\thco}{$^{13}$CO\xspace}

\newcommand{\kks}{${\rm K\,km\,s^{-1}}$\xspace}

\newcommand{\kms}{${\rm km\,s^{-1}}$\xspace}
\newcommand{\htwo}{H$_{2}$\xspace}
\newcommand{\msol}{M$_{\odot}$\xspace}

\newcommand{\hi}{{\sc H\,i}\xspace}

\newcommand{\hibold}{{\bf H\,{\sc i}}\xspace}

\defcitealias{Koch2018MNRAS}{K18}

\graphicspath{{./}{figures/}} 

\title[Atomic and Molecular Line Widths in M33]{Relationship between the Line Width of the Atomic and Molecular ISM in M33}

\author[Koch et al.]{Eric W. Koch$^{1}$\thanks{E-mail:
koch.eric.w@gmail.com (EWK); rosolowsky@ualberta.ca (EWR)}, Erik W. Rosolowsky$^{1}$, Andreas Schruba$^{2}$, Adam Leroy$^{3}$,
\newauthor Amanda Kepley$^{4}$, Jonathan Braine$^{5}$, Julianne Dalcanton$^{6}$, Megan C. Johnson$^{7}$\\
$^{1}$University of Alberta, Department of Physics, 4-183 CCIS, Edmonton AB T6G 2E1, Canada \\
$^{2}$Max-Planck-Institut f\"ur extraterrestrische Physik, Giessenbachstra\ss e 1, D-85748 Garching, Germany\\
$^{3}$The Ohio State University, Department of Astronomy, 140 West 18th Avenue, Columbus, OH 43210, USA\\
$^{4}$National Radio Astronomy Observatory, 520 Edgemont Road, Charlottesville, VA 22903-2475, USA\\
$^{5}$Laboratoire d'Astrophysique de Bordeaux, Univ. Bordeaux, CNRS, B18N, all\'ee Geoffroy Saint-Hilaire, 33615 Pessac, France\\
$^{6}$Department of Astronomy, Box 351580 University of Washington, Seattle, WA 98195\\
$^{7}$United States Naval Observatory, 3450 Massachusetts Ave NW, Washington, D.C., 20392, USA}

\begin{document}

\date{Draft date: \today}

\pagerange{\pageref{firstpage}--\pageref{lastpage}} \pubyear{2018}

\maketitle

\label{firstpage}

\begin{abstract}
We investigate how the spectral properties of atomic (\hi) and molecular (\htwo) gas, traced by \cotwoone, are related in M33 on $80$~pc scales.  We find the \hi and \cotwoone velocity at peak intensity to be highly correlated, consistent with previous studies.  By stacking spectra aligned to the velocity of \hi peak intensity, we find that the \co\ line width ($\sigma_{\rm HWHM}=4.6\pm0.9$~\kms; $\sigma_{\rm HWHM}$ is the effective Gaussian width) is consistently smaller than the \hi line width ($\sigma_{\rm HWHM}=6.6\pm0.1$~\kms), with a ratio of ${\sim}0.7$, in agreement with \citet{Druard2014A&A...567A.118D}. The ratio of the line widths remains less than unity when the data are smoothed to a coarser spatial resolution. In other nearby galaxies, this line width ratio is close to unity which has been used as evidence for a thick, diffuse molecular disk that is distinct from the thin molecular disk dominated by molecular clouds.  The smaller line width ratio found here suggests that M33 has a marginal thick molecular disk.
From modelling individual lines-of-sight, we recover a strong correlation between \hi and \co line widths when only the \hi located closest to the \co component is considered.  The median line width ratio of the line-of-sight line widths is $0.56\pm0.01$.  There is substantial scatter in the \hi--\cotwoone line width relation, larger than the uncertainties, that results from regional variations on $<500$~pc scales, and there is no significant trend in the line widths, or their ratios, with galactocentric radius.  These regional line width variations may be a useful probe of changes in the local cloud environment or the evolutionary state of molecular clouds.
\end{abstract}


\section{Introduction}
\label{sec:intro}

Across large samples of nearby galaxies, several studies show a tight correlation between the surface density of molecular (\htwo) gas and star formation rate (SFR) surface density \citep{Kennicutt1998ApJ...498..541K,Leroy2008AJ....136.2782L,Bigiel2008AJ....136.2846B,Kennicutt2011PASP..123.1347K}, and a lack of correlation with the atomic (\hi) gas surface density \citep{Bigiel2008AJ....136.2846B,Schruba2011AJ....142...37S}.  This result shows that star formation is primarily coupled to the molecular gas, rather than the total (\hi\ + \htwo) gas component.

A critical, potentially rate-limiting, step in the star formation process is then the formation of molecular gas.  Several mechanisms have been proposed that lead to conditions where molecular gas can readily form \citep{Dobbs2014prpl.conf....3D}.  These mechanisms for forming the molecular interstellar medium (ISM) are predicted to act over scales ranging from individual molecular clouds to galactic scales.  Recent star formation models have sought to predict the atomic-to-molecular gas fraction from the local environment properties \citep{Blitz2006ApJ...650..933B,Krumholz2009ApJ...693..216K,Ostriker2010ApJ...721..975O,Krumholz2013MNRAS.436.2747K,Sternberg2014ApJ...790...10S,Bialy2017ApJ...843...92B} and recover observed properties to within a factor of a few \citep{Bolatto2011ApJ...741...12B,Jameson2016ApJ...825...12J,Schruba2018ApJ...862..110S}.

To observe signatures of the molecular ISM, we require observations that resolve giant molecular cloud (GMC) scales ($<100$~pc) in both the atomic and molecular gas.  Only within the Local Group can current 21-cm telescopes resolve GMC scales, making studies of M33, M31, and the Magellanic Cloud critical for understanding how the molecular ISM forms.  In this paper, we use \hi and \cotwoone observations of M33 with a resolution of $80$~pc to study the spectral properties of the atomic and molecular ISM.

Previous high-resolution studies of \hi and \co, used as a tracer of \htwo, in the Local Group have identified spectral-line properties that are correlated between these tracers.  \citet{Wong2009ApJ...696..370W} and \citet{Fukui2009ApJ...705..144F} compared the \hi to \co properties in the Large Magellanic Cloud (LMC) on $40$~pc scales.  They found that \hi and \co spectral properties are correlated, with a close relationship between the velocities at peak intensity and a suggestive correlation between the \hi and \co line widths.  However, they also found that the \hi temperature and column density are poor predictors for the detection of \co, suggesting that a significant amount of \hi emission arises from atomic gas not associated with the molecular gas.

On larger scales ($>100$~pc) where individual clouds are unresolved, several studies have found evidence of a large-scale molecular component, possibly unassociated with \co emission from GMCs on small scales. \citet{GarciaBurillo1992A&A...266...21G} found \co emission $\sim1$~kpc from the plane of the disk in the edge-on galaxy NGC~891, providing direct evidence for a ``molecular halo.''  More recently, \citet{Pety2013ApJ...779...43P} find evidence for a diffuse molecular disk based on interferometric data ($\sim50$~pc resolution) recovering only $\sim50\%$ of the flux from single-dish data.  They suggest that the remaining emission is filtered out by the interferometer and must be from larger scales.  Using a similar comparison between interferometric and single-dish data, \citet{CalduPrimo2015AJ....149...76C} and \citet{CalduPrimo2016AJ....151...34C} identify a wide velocity component in the \co that is only recovered in single-dish data on scales $>500$~pc.

There is also growing evidence for a significant diffuse molecular component in the Milky Way. \citet{Dame1994ApJ...436L.173D} find excess \co emission in the line wings that may be similar to the wide velocity components in nearby galaxies \citep{CalduPrimo2015AJ....149...76C,CalduPrimo2016AJ....151...34C}. \citet{RomanDuval2016ApJ...818..144R} find $25\%$ of the Milky Way molecular gas mass is in diffuse $^{12}$CO emission that is extended perpendicular to the Galactic plane beyond the $^{12}$CO emission where denser gas is detected.

Spectral analyses have found connections between the \hi with the bright dense and faint diffuse \co components.  The different \co components are highlighted through different analyses, with individual lines-of-sight primarily tracing the bright \co emission, while analyses that study an ensemble of spectra through stacking recover the faint \co emission.  Comparing these analyses shows that the properties of the bright and faint \co emission differ. \citet{Fukui2009ApJ...705..144F} find \co line widths in the LMC on $40$~pc scales that are $\sim30\%$ of the \hi line widths along the same lines-of-sight.  Similar ratios between the \co and \hi are found by \citet{Wilson2011MNRAS.410.1409W} for 12 nearby galaxies on scales from $\sim200\mbox{--}1200$~pc, though the \hi line widths are estimated at a different resolution from the \co.  On similar scales ($\sim200\mbox{--}700$~pc), with matched resolution between the \hi and \co, \citet{Mogotsi2016AJ....151...15M} found that the \co line widths ($\sigma_{\rm CO}=7.3\pm1.7$~\kms) are consistently narrower than the \hi ($\sigma_{\rm HI}=11.7\pm2.3$~\kms) for a number of nearby galaxies. The average ratio of $\sim0.6$ between the line widths is much larger than the ratio from \citet{Fukui2009ApJ...705..144F} on smaller scales ($\sim0.3$).

Stacking analyses consistently have broader \co line widths than those from individual spectra.  \citet{Combes1997A&A...326..554C} found comparable \hi and \co line widths in two nearby face-on galaxies ($i<12\degree$).  They suggested that the \hi and \co emission trace a common, well-mixed kinematic component that differs only in the phase of the gas.  Using the same data as \citet{Mogotsi2016AJ....151...15M}, \citet{Caldu2013AJ....146..150C} also found similar line widths between the \hi and \co  ($\sigma_{\rm CO}=12.0\pm3.9$~\kms and $\sigma_{\rm HI}=11.9\pm3.1$~\kms) for a number of nearby galaxies. \citet{Caldu2013AJ....146..150C} concluded that the wide \co component arises from a faint, large-scale molecular component that is too faint to be detected in individual lines-of-sight.  However, a stacked spectrum is broadened due to scatter in the line centre \citep{Koch2018MNRAS}, particularly when \hi velocities are used to align the \co spectra \citep{Schruba2011AJ....142...37S,Caldu2013AJ....146..150C}.  Characterizing methodological sources of line broadening is critical for understanding the spectral properties of the diffuse molecular component.

In M33, there are differing results regarding a diffuse molecular component. \citet{Wilson1990ApJ...363..435W} inferred the presence of diffuse molecular gas from interferometric data recovering $\sim40\%$ of the flux from single-dish observations. \citet{Wilson1994ApJ...432..148W} supported this conclusion by demonstrating that the high $^{12}$CO to $^{13}$CO line ratio does not result from different filling factors between the two lines. Later, \citet{Combes2012A&A...539A..67C} found a non-zero spatial power-spectrum index on kpc scales and suggested that it arises from a large-scale \co component.

\citet{Roso2003ApJ...599..258R} and \citet{Rosolowsky2007ApJ...661..830R} also found additional \co emission that did not arise from GMCs, similar to \citet{Wilson1990ApJ...363..435W}. However, \citet{Roso2003ApJ...599..258R} localized $90\%$ of the diffuse emission to within $100$~pc of a GMC and suggested that this diffuse emission is from a population of small, unresolved molecular clouds that are too faint for their interferometric observations to detect.

These previous results in M33 and other nearby galaxies suggest that detailed studies of molecular clouds and their local environments may need to account for the presence of diffuse \co emission or bright \hi emission along the line-of-sight that is unrelated to the molecular cloud.  In this paper, we characterize the relationship between the spectral properties of \hi and \co in M33 on $80$~pc scales by stacking spectra and modelling individual lines-of-sight. We then critically compare these two different analyses, constraining how methodological line broadening and unrelated \hi or \co emission affects the properties of stacked spectra.  M33 is an ideal system for this comparison as we can connect studies of \hi and \co performed on larger scales ($>100$~pc) to those on small scales ($<50$~pc).

M33's flocculent morphology also lies between the nearby galaxies in previous studies, with a sample of more massive spiral galaxies in the lower-resolution studies and the irregular morphology of the LMC observed at higher-resolution.  The \hi in M33 is an ideal tracer of the flocculent spiral structure.  The bright \hi is aligned in filaments, similar to the ``high-brightness network'' identified in other galaxies \citep{Braun1997ApJ...484..637B}.

We compare the atomic and molecular ISM using \hi observations obtained with the Karl G.\ Jansky Very Large Array (VLA) by \citet[][hereafter \citetalias{Koch2018MNRAS}]{Koch2018MNRAS} and the \cotwoone data from the IRAM \mbox{30-m} by \citet{Druard2014A&A...567A.118D}, as described in \S\ref{sec:observations}.  The \hi data have a beam size of $20$\arcsec, corresponding to physical scales of ${\sim}80$~pc at the distance of M33 \citep[840 kpc;][]{Freedman2001ApJ...553...47F}.  Our study builds on work by \citet{Fukui2009ApJ...705..144F} and \citet{Druard2014A&A...567A.118D} by utilizing improved \mbox{21-cm} \hi observations and new techniques for identifying spectral relationships.  We focus on comparing M33's \hi and \co distributions along the same lines-of-sight, where we explore the difference in velocity where the \hi and \co intensity peaks (\S\ref{sub:peak_velocity_relation}), how the line widths of stacked line profiles compare to those measured at lower resolutions (\S\ref{sub:total_profiles}), and the distribution of \hi and \co line widths from fitting individual spectra (\S\ref{sub:hi_line_widths_associated_with_co}).  We then compare the properties from these two analyses and discern where sources of discrepancy arise (\S\ref{sub:comparing_stacking_and_los}).  Our results show that M33 does not have a significant diffuse molecular disk.  We discuss this result and compare to previous findings in \S\ref{sec:no_thick_molecular_disk_in_m33_}.

\section{Observations}
\label{sec:observations}

\subsection{\hibold VLA \& GBT} 
\label{sub:vla}

\begin{figure*}
\includegraphics[width=\textwidth]{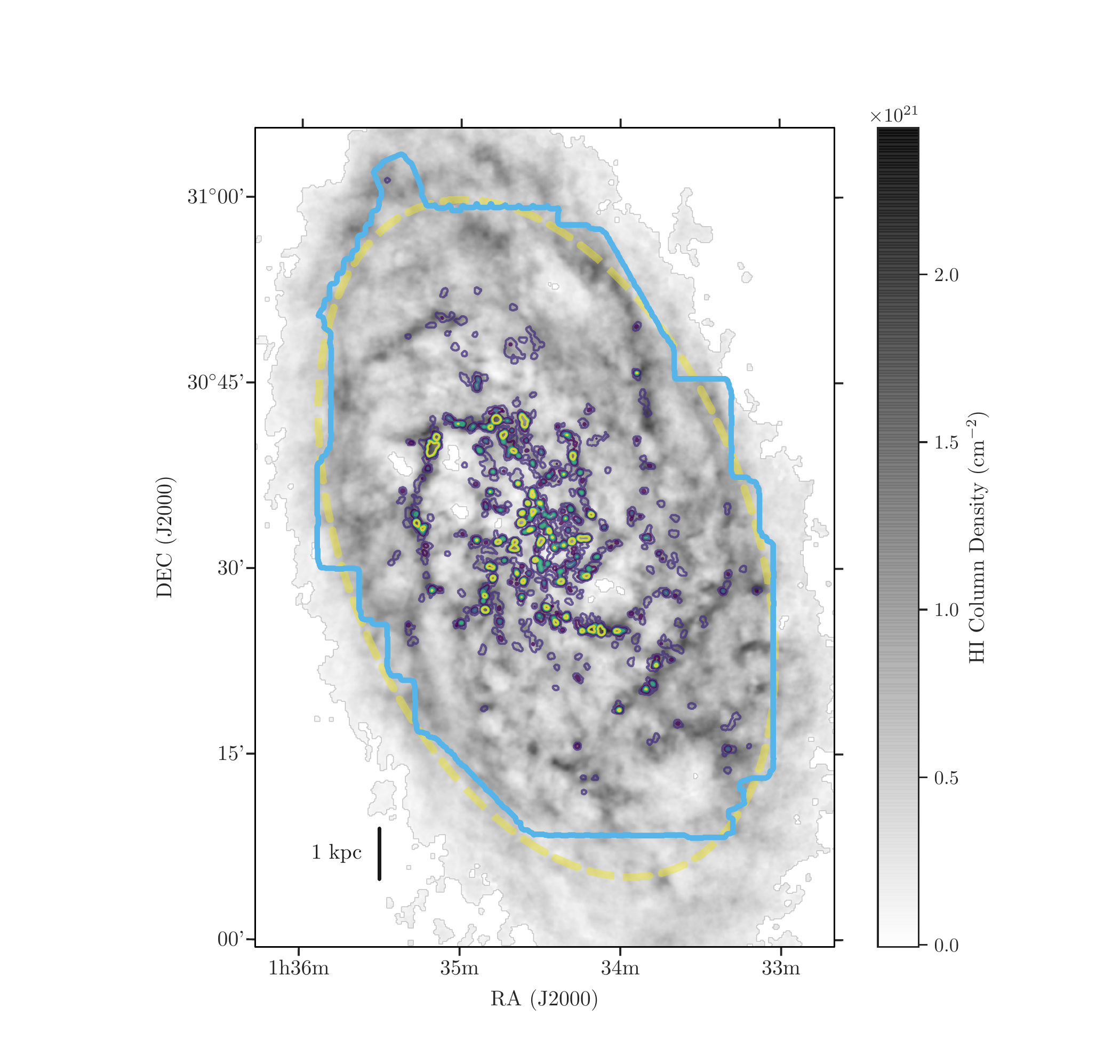}
\caption{\label{fig:hi_co_coldens} \hi and \cotwoone (contours) column density maps at a resolution of $\sim80$~pc ($20\arcsec$).  The \hi column density assumes optically thin emission and is corrected for inclination \citepalias{Koch2018MNRAS}.  The \cotwoone contour levels (from blue to yellow) indicate surface densities of 900, 1400, 1900, and 2400~\kks.
The light-blue line indicates the extent of the \cotwoone data, and the dashed yellow line shows the $R_{\rm gal}=7$~kpc galactocentric radius.  Qualitatively, the \cotwoone emission tends to be located with bright \hi.}
\end{figure*}

We utilize the \hi observations presented in \citetalias{Koch2018MNRAS} and provide a short summary of the observations here.  Figure \ref{fig:hi_co_coldens} shows the \hi integrated intensity map.  The observations were taken with the VLA using a 13-point mosaic to cover the inner 12~kpc of M33.  The data were imaged with \mbox{CASA~4.4} using natural weighing and deconvolved until the peak residual reached 3.8 mJy~beam$^{-1}$ ($7.1$~K) per channel, which is about $2.5$ times the noise level in the data.  The resulting data cube has a beam size of $20\arcsec\times18\arcsec$, a spectral resolution of $0.2$~\kms, and a $1\mbox{-}\sigma$ noise level of $2.8$~K per channel.  This spectral resolution is a factor of $\sim13$ finer than the \cotwoone data (\S\ref{sub:iram_30_m}), leading to significantly less uncertainty in the velocity at peak intensity in the \hi compared to the \cotwoone.

We combine the VLA data with GBT observations by \citet{Lockman2012AJ....144...52L} to include short-spacing information\footnote{Described in Appendix A of \citetalias{Koch2018MNRAS}}. We feather the data sets together using the {\sc uvcombine} package\footnote{\url{https://github.com/radio-astro-tools/uvcombine}}, which implements the same feathering procedure as CASA.  Thus the \hi data used in this work provide a full account of the \hi emission down to ${\sim}80$~pc scales.

\subsection{\cotwoone IRAM 30-m} 
\label{sub:iram_30_m}

We use the \cotwoone data from the IRAM-30m telescope presented by \citet{Druard2014A&A...567A.118D}. Figure \ref{fig:hi_co_coldens} shows the region covered by these observations, along with the zeroth moment contours.  A full description of the data and reduction process can be found in their \S2; a brief summary is provided here. Portions of the map were previously presented by \citet{Gardan2007A&A...473...91G}, \citet{Gratier2010A&A...522A...3G}, and \citet{Gratier2012A&A...542A.108G}. The data have an angular resolution of $12\arcsec$, corresponding to a physical resolution of ${\sim}48$~pc, and a spectral resolution of $2.6$~\kms.  Because IRAM 30-m is a single dish telescope, the data are sensitive to all spatial scales above the beam size and does not require the feathering step used with the \hi (\S\ref{sub:vla}).

The \cotwoone cube is a combination of many observations that leads to spatial variations in the noise.  The rms noise level differs by a factor of a few in the inner $\sim7$ kpc of M33's disk \citep[see Figure~6 in][] {Druard2014A&A...567A.118D}. We adopt the same beam efficiency of $0.56 / 0.92 = 0.61$ from \citet{Druard2014A&A...567A.118D} for converting to the main beam temperature. The average noise per channel is $33.3$~mK in units of the main beam temperature.  Since we focus only on the line shape properties, we do not require a conversion factor to the \htwo column density in this paper.

The spectral channels are moderately correlated due to the spectral response function of the instrument.  Along with broadening due to finite channel widths, the spectral response function correlates nearby channels and broadens the spectra.  This broadening can be accounted for by modelling the known spectral response function and accounting for the channel width \citep{Koch2018-linewidths}.  Adopting the correlation coefficient of $r=0.26$ determined by \citet{Sun2018ApJ...860..172S} for these data, and using the empirical relation from \citet{Leroy2016ApJ...831...16L}, we approximate the spectral response function as a three-element Hanning-like kernel with a shape of $[k,\ 1 - 2k,\ k]$, where $k=0.11$ is the channel coupling.  The use of the spectral response function in spectral fitting is described further in \S\ref{subsub:fitting_individual_spectra}.

Throughout this paper, we use a spatially-matched version of these \cotwoone data convolved to have the same beam size as the \hi data .  The data are spatially-convolved and reprojected to match the \hi data, which lowers the average noise per channel to $16.0$~mK.  The spectral dimension is not changed.  We create a signal mask for the data by searching for connected regions in the data with a minimum intensity of 2-$\sigma$ that contain a peak of at least 4-$\sigma$.  Each region in the mask must be continuous across three channels and have a spatial size larger than the full-width-half-maximum of the beam.


\section{\hibold-CO Spectral Association}
\label{sec:hi_and_co_spectral}

We examine the relation between \hi and \cotwoone spectra using three comparisons: (i) the distribution of peak velocity offsets; (ii) the width and line wing excess, and shape parameters of stacked profiles; and (iii) the line widths of both tracers from a limited Gaussian decomposition of the \hi associated with \cotwoone emission.  Unless otherwise specified, the line width refers to the Gaussian standard deviation ($\sigma$) and not the full-width-half-maximum (${\rm FWHM} = 2\sqrt{2 \ {\ln} \ 2} \sigma$).

\subsection{Peak Velocity Relation} 
\label{sub:peak_velocity_relation}

\begin{figure}
\includegraphics[width=0.5\textwidth]{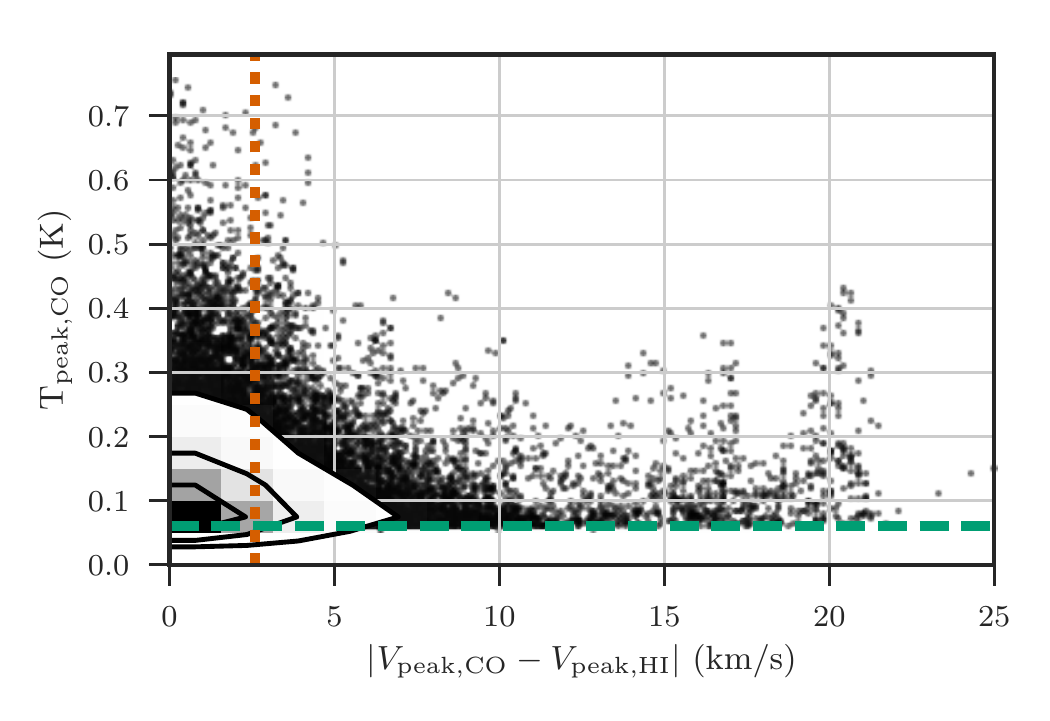}
\caption{\label{fig:co_peak_vel_offset} Distribution of the peak \cotwoone brightness (with ${\rm S/N}>3$) versus the absolute difference in the \hi and \cotwoone peak velocities. The shaded region and contours indicate the regions containing the 1, 2, and 3-$\sigma$ limits of the distribution of points.  Individual points show outliers beyond $3\sigma$.  The dashed horizontal line is the $3\sigma$ rms noise cut-off in the \cotwoone data imposed to avoid spurious outliers in the velocity difference.  The dotted vertical line is the \co channel width of $2.6$~\kms.  The \co velocities are preferentially located at or near the \hi velocities. However, there remains a number of high S/N outliers with a large velocity difference.  These outliers occur when the CO emission is associated with a different HI component than the brightest HI peak.
}
\end{figure}

We first determine the spectral relation between the \hi and \cotwoone by comparing the velocity of the peak temperatures along the same lines-of-sight.  We refer to this velocity as the ``peak velocity.''  Figure~\ref{fig:co_peak_vel_offset} compares the absolute peak velocity difference between \hi and \cotwoone versus the peak \co temperature.  Most lines-of-sight have peak velocities consistent between the \hi and \cotwoone.  The standard deviation of the velocity difference, after removing severe outliers with differences of $>10$~\kms, is $2.7$~\kms.  This is similar to the $2.6$~\kms channel width of the \cotwoone data, suggesting that the peak velocities are typically consistent within the resolution of the \cotwoone data.  Since the peak velocities are defined at the centre of the velocity channel at peak intensity, recovering a scatter in the peak velocity difference of $\sim\pm1$ channel is reasonable.  The much narrower \hi channel width ($0.21$ \kms) accounts for significantly less scatter than the \cotwoone channel width.

Previous \hi-\co\ studies find a similar correlation between the peak velocities of these tracers and have used the \hi to infer the peak velocity of \cotwoone with the goal of detecting faint \co  \citep{Schruba2011AJ....142...37S,Caldu2013AJ....146..150C}.  The brightest \cotwoone peak intensities tend to have smaller velocity differences between the \hi and \co, also consistent with the relation found on $40$~pc scales in the LMC by \citet{Wong2009ApJ...696..370W}.

The distribution in Figure~\ref{fig:co_peak_vel_offset} has several outliers with velocity differences of $>10$~\kms, far larger than what would be expected from a Gaussian distribution with a width of $2.7$~\kms.  These outliers account for $3\%$ of the lines of sight and result from locations where the \hi spectrum has multiple components and the \cotwoone peak is not associated with the brightest \hi peak \citep{Gratier2010A&A...522A...3G}.  In these cases, the \cotwoone peaks are well-correlated with the peak of the fainter \hi component (\S\ref{sub:hi_line_widths_associated_with_co}).  This result is important when stacking spectra (\S\ref{sub:total_profiles}) aligned with respect to the peak \hi temperature.  When the \cotwoone peak is not associated with the brightest \hi peak, the \cotwoone stacked profile will be broadened and could potentially be asymmetric if the \co peaks are preferentially blue- or red-shifted from the \hi component.  We explore these effects in \S\ref{sub:comparing_stacking_and_los}.

We conclude that the \hi peak velocity can nearly always be used to infer the \cotwoone peak velocity.


\subsection{Stacking Analysis} 
\label{sub:total_profiles}

By stacking a large number of spectra aligned to a common velocity, we can examine a high signal-to-noise (S/N) average spectrum of each tracer.  Since the signal will add coherently, while the noise will add incoherently, the stacked profiles are ideal for identifying faint emission that is otherwise not detectable in individual lines-of-sight \citep{Schruba2011AJ....142...37S}.  These high S/N spectra can be used to compare the line profile properties of the \hi and \cotwoone.

We examine stacked profiles of \hi and \cotwoone aligned with respect to the \hi peak velocity since the \hi is detected towards nearly every line-of-sight, the \hi peak velocity describes the peak \cotwoone velocity well (\S\ref{sub:peak_velocity_relation}), and the velocity resolution of the \hi data is much higher than the \cotwoone data.  We align the spectra by shifting them; we Fourier transform the data, apply a phase shift, and transform back.  This procedure preserves the signal shape and noise properties when shifting by a fraction of the channel size\footnote{Implemented in spectral-cube (\url{https://spectral-cube.readthedocs.io})}.  The channel size is a particular issue for the \cotwoone data, since the channels have a width of $2.6$~\kms and the signal in some spectra only spans ${\sim}5$ channels.

\begin{figure}
\includegraphics[width=0.5\textwidth]{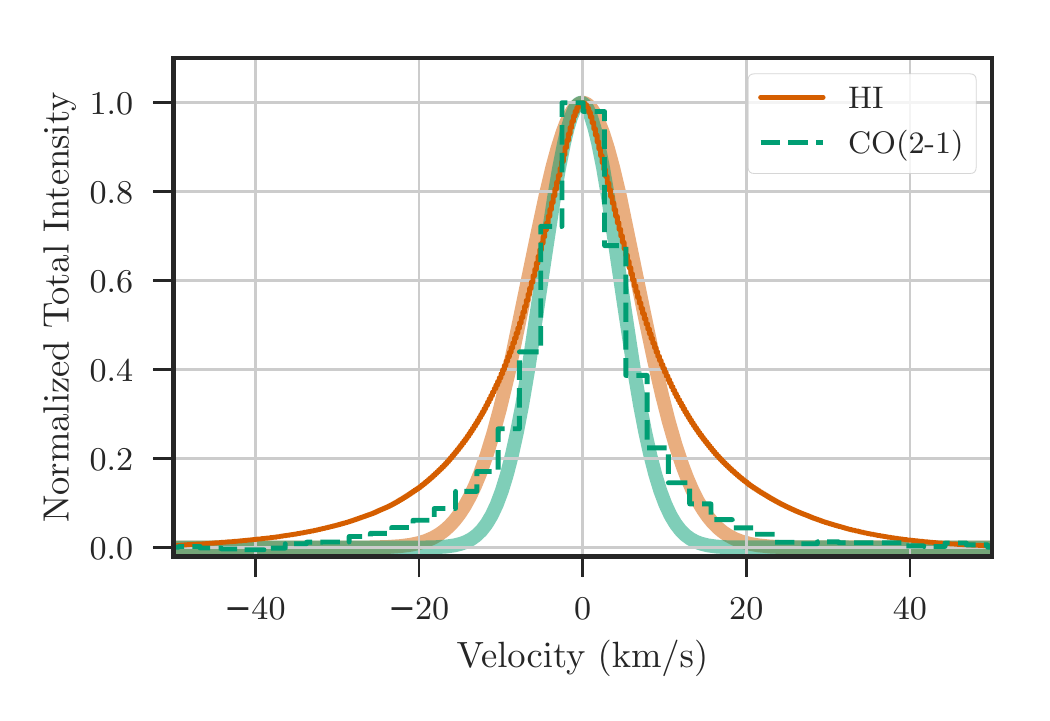}
\caption{\label{fig:total_profs_co_hi} The \hi (orange solid) and \cotwoone (green dashed) stacked profiles shifted with respect to the \hi peak velocity. The thick faint lines show the Gaussian model for each tracer.  The \hi stacked profile is wider and has a larger line wing excess than the \cotwoone stacked profile.}
\end{figure}

Figure~\ref{fig:total_profs_co_hi} shows the stacked profiles, where spectra within a radius of $7$~kpc are included.  The \hi stacked profile has a kurtosis excess relative to a Gaussian, with enhanced tails and a steep peak.  These properties of \hi stacked profiles are extensively discussed in \citetalias{Koch2018MNRAS}.  The \cotwoone stacked profile has a qualitatively similar shape but is narrower than the \hi profile and has a smaller line wing excess.

As in \citetalias{Koch2018MNRAS}, we model the profiles based on the half-width-half-maximum (${\rm HWHM} = {\rm FWHM}/2$) approach from \citet{Stilp2013ApJ...765..136S}.  The model and parameter definitions are fully described in \citetalias{Koch2018MNRAS}; we provide a brief overview here.  The HWHM model assumes that the central peak of the profiles can be described by a Gaussian profile whose FWHM matches the profile's FWHM, which is well-constrained in the limit of high S/N.  This model sets the Gaussian standard deviation ($\sigma_{\rm HWHM}={\rm HWHM}/\sqrt{2\, {\rm ln}\, 2}$) and central velocity ($v_{\rm peak}$) of this Gaussian, which we refer to as $G(v)$.

The following parameters describe how the observed profile, $S(v)$, compares to the Gaussian model.

The line wing excess expresses the fractional excess relative to the Gaussian outside of the FWHM:
\begin{equation}
  \label{eq:f_wings}
  f_{\rm wings} = \frac{\sum\limits_{|v|>{\rm HWHM}} \left[S(v) - G(v)\right]}{\sum\limits_{v} S(v)}~.
\end{equation}
This excess in the line wings can also be used to find the ``width'' of the wings using a form equivalent to the second moment of a Gaussian:
\begin{equation}
  \label{eq:sigma_wings}
  \sigma_{\rm wings}^2 = \frac{\sum\limits_{|v|>{\rm HWHM}} \left[S(v) - G(v)\right]v^2}{\sum\limits_{|v|>{\rm HWHM}} \left[S(v) - G(v)\right]}~.
\end{equation}
Since the line wing excess will not be close to Gaussian in shape, $\sigma_{\rm wings}$ does not have a clear connection to a Gaussian width.

The asymmetry of a stacked profile is defined as the difference in total flux at velocities greater than and less than $v_{\rm peak}$, normalized by the total flux:
\begin{equation}
  \label{eq:aymm_hwhm}
  a = \frac{\sum\limits_{v > v_{\rm peak}} S(v) - \sum\limits_{v < v_{\rm peak}} S(v)}{\sum S(v)}~.
\end{equation}
This makes $a$ analogous to the skewness of the profile.

The shape of the peak is described by $\kappa$, defined as the fractional difference between the central peak and the Gaussian model within the FWHM:
\begin{equation}
  \label{eq:kappa_hwhm}
  \kappa = \frac{\sum\limits_{|v|<{\rm HWHM}} \left[ S(v) - G(v) \right]}{\sum\limits_{|v|<{\rm HWHM}} G(v)}~.
\end{equation}
This describes the kurtosis of the profile peak, where $\kappa > 0$ is a profile more peaked than a Gaussian and $\kappa < 0$ is flatter than a Gaussian.  We note that the kurtosis typically describes the line wing structure, however since these regions are excluded in our definition, $\kappa$ describes the shape of the peak.

Since we adopt a semi-parametric model for the stacked profiles, deriving parameter uncertainties also requires a non-parametric approach.  We use a bootstrap approach presented in \citetalias{Koch2018MNRAS} to account for the two significant sources of uncertainty:
\begin{enumerate}
    \item {\it Uncertainty from the data} -- These uncertainties come directly from the noise in the data.  In each channel, the uncertainty is $\sigma_{\rm RMS} / \sqrt{N_{\rm spec}}$, where $N_{\rm spec}$ is the number of spectra included in the stacked spectrum.  We account for this uncertainty by resampling the values in the stacked spectrum by drawing from a normal distribution centered on the original value with a standard deviation equal to the noise.  We then calculate the parameters from the resampled stacked spectrum at each iteration in the bootstrap.
    \item {\it Uncertainty due to finite channel width} --  The finite channel width introduces uncertainty in the location of the peak velocity when not explicitly modelled for with an analytic model \citep{Koch2018-linewidths}.  Since the HWHM model is a semi-parametric model that does not account for finite channel width, we need to adopt an uncertainty for the peak velocity and the inferred line width.  We use the HWHM model on very high S/N stacked spectra and assume that the true peak velocity is known to be within the channel of peak intensity.  To create an equivalent Gaussian standard deviation\footnote{In \citetalias{Koch2018MNRAS}, we used $\Delta v$ as the uncertainty. Since the \hi channels are much narrower than $\sigma_{\rm HWHM}$, this change has little effect on the \hi uncertainties.} for the uncertainty, we scale the rectangular area in the peak channel to the fraction of the area under a Gaussian within $\pm1\mbox{--}\sigma$, which gives $\sigma_{\rm v_{\rm peak}}= 0.34 \Delta v$\footnote{$A\, (2\,\sigma_{\rm v_{\rm peak}})= A \, (0.68 \Delta v)$, where $A$ is the value in the stacked spectrum and $0.68$ is the fraction of the area when integrating a Gaussian from $-1\sigma$ to $+1\sigma$.}. The HWHM model estimates the width $\sigma_{\rm HWHM}$ based on $v_{\rm peak}$ and thus we adopt the same uncertainty for both parameters.   To estimate the uncertainties on the other parameters in the HWHM model, we sample new values of $v_{\rm peak}$ and $\sigma_{\rm HWHM}$ from normal distributions with standard deviations of $0.34 \Delta v$ in each bootstrap iteration.  These two parameters set the Gaussian shape used to derived the parameters in Equations \ref{eq:f_wings}--\ref{eq:kappa_hwhm}.
\end{enumerate}

Based on the bootstrap sampling used in these two steps, we estimate the uncertainties on the remaining HWHM model parameters.  Since the stacked profiles have high S/N, the parameter uncertainties are dominated by the uncertainty in $v_{\rm peak}$ and $\sigma_{\rm HWHM}$, and since the \cotwoone channel size is much larger than the \hi ($2.6$ versus $0.2$ \kms), the \cotwoone uncertainties are much larger.

Table~\ref{tab:global_profiles} provides the parameter values and uncertainties from the HWHM model for the stacked profiles in Figure~\ref{fig:total_profs_co_hi}.  The \cotwoone line width is $4.6\pm0.9$ \kms, which is $70\%$ of the \hi width of $6.6\pm0.1$ \kms.

Using the same \cotwoone data at the original $12\arcsec$ resolution, \citet{Druard2014A&A...567A.118D} create \cotwoone stacked profiles and fit a single Gaussian component to the profile.  They find a line width of $\sigma=5.3\pm0.2$ \kms, which is $0.7$ \kms larger than our measurement using the HWHM method.  This discrepancy results from the different modelling approaches used; fitting our \cotwoone stacked profile with a single Gaussian component gives a line width of $5.4\pm0.9$~\kms, consistent with \citet{Druard2014A&A...567A.118D}, because the fit is influenced by the line wings.

The profiles are consistent with the same $v_{\rm peak}$, as is expected based on the strong agreement between the peak velocities (\S\ref{sub:peak_velocity_relation}). The scatter in the peak velocity difference will primarily {\it broaden} the spectrum, rather than create an offset in $v_{\rm peak}$.  We test the importance of this source of broadening in \S\ref{sub:comparing_stacking_and_los}, but note that the non-Gaussian shape of the stacked profiles makes correcting for this broadening non-trivial.  Because of this, we do not apply a correction factor to $\sigma_{\rm HWHM}$ to account for the spectral response function and channel width since the stacked profiles have a non-Gaussian shape.

\begin{table}
\caption{\label{tab:global_profiles} HWHM model parameters for the \hi and \cotwoone stacked profiles for $R_{\rm gal} < 7$~kpc at different spatial resolutions. The uncertainties are propagated assuming an uncertainty of half the channel width and the uncertainty of each point in the spectrum is the standard deviation of values within that channel scaled by the square-root of the number of beams.}
\centering
\input{tables/total_hwhm_fits_peakonly.tex}
\end{table}

The \hi profile is more non-Gaussian in shape than the \cotwoone.  The \hi profile has a larger line wing excess and a non-Gaussian peak ($\kappa < 0$), consistent with the stacked profiles in \citetalias{Koch2018MNRAS}.

The large uncertainties on the \cotwoone shape parameters make most not significant at the 1-$\sigma$ level, or are consistent within 1-$\sigma$ of the \hi shape parameters.  Within the uncertainty, the \cotwoone stacked profiles are symmetric about the peak ($a=0$) and have a Gaussian-shaped profile within the HWHM ($\kappa=0$).  The only significant \cotwoone shape parameter is the line wing excess $f_{\rm wings}$, which is non-zero at the 2-$\sigma$ level and consistent with the \hi $f_{\rm wings}$ within 1-$\sigma$.  However, there are additional systematics that may contribute to the \cotwoone $f_{\rm wings}$, including broadening from the distribution of the peak \hi and \co velocities (\S\ref{sub:peak_velocity_relation}) and the IRAM \mbox{30-m} error beam pickup \citep{Druard2014A&A...567A.118D}.  We discuss the former contribution in more detail in \S\ref{sub:comparing_stacking_and_los}.  \citet{Druard2014A&A...567A.118D} estimate that the error-beam pickup contributes $2.5\times10^6$ \msol, or $1.1\times10^4$ \kks, using their conversion factor $X_{\rm CO} = 4 \times 10^{20}$ ${\rm cm}^{-2} / ({\rm K\ km\ s}^{-1})$ and a brightness temperature ratio of $0.8$ between the $J=2\mbox{-}1$ and $J=1\mbox{-}0$\, \co transitions.  The error beam flux may then contribute up to $45\%$ of the $2.4\times10^{4}$ \kks line wing excess.  We further assess whether the error beam flux contributes to the line wings in \S\ref{subsub:radial_stacked_profiles}.

Assuming that the error beam does contribute $45\%$ of the \co line wing excess, M33 appears to exhibit weaker line wings compared to those measured in M31 by \citet{CalduPrimo2016AJ....151...34C}.  They characterized the \co stacked profiles with a Gaussian model and found that single-dish CO observations in M31 are best fit by two Gaussian components.  Their wide Gaussian component would be related to the line wing excess, i.e., a large $f_{\rm wings}$, in our formalism\footnote{A relation between $f_{\rm wings}$ and the wide Gaussian component would be a function of the amplitudes and widths of both Gaussian components, and the $\sigma_{\rm HWHM}$ used here.}.  For the sake of comparison with \citet{CalduPrimo2016AJ....151...34C}, we fit a two-Gaussian component to the \cotwoone stacked profile and find line widths of $3.8\pm0.9$ \kms and $10.9\pm0.9$ \kms for the narrow and wide Gaussian components, respectively.  The narrow line width is similar to the $3.2\pm0.2$ \kms found by \citet{CalduPrimo2016AJ....151...34C}, however their wide component is much narrower, with a width of $6.1\pm0.6$ \kms.

\subsubsection{Radial Stacked Profiles}
\label{subsub:radial_stacked_profiles}

We explore trends with galactocentric radius by creating stacked profiles within radial bins of $500$~pc widths out to a maximal radius of $7$~kpc, matching the coverage of the \cotwoone map.  We use a position angle of $201.1\degree$ and inclination angle of $55.1\degree$ for M33's orientation, based on the \hi kinematics from \citetalias{Koch2018MNRAS}.  The radial stacking uses the same procedure for the \hi profiles as in \citetalias{Koch2018MNRAS}, but with $500$~pc radial bins instead of $100$~pc due to the smaller filling fraction of \cotwoone detections relative to the \hi. The stacked profiles are modeled with the same HWHM model described above.

\begin{figure}
\includegraphics[width=0.5\textwidth]{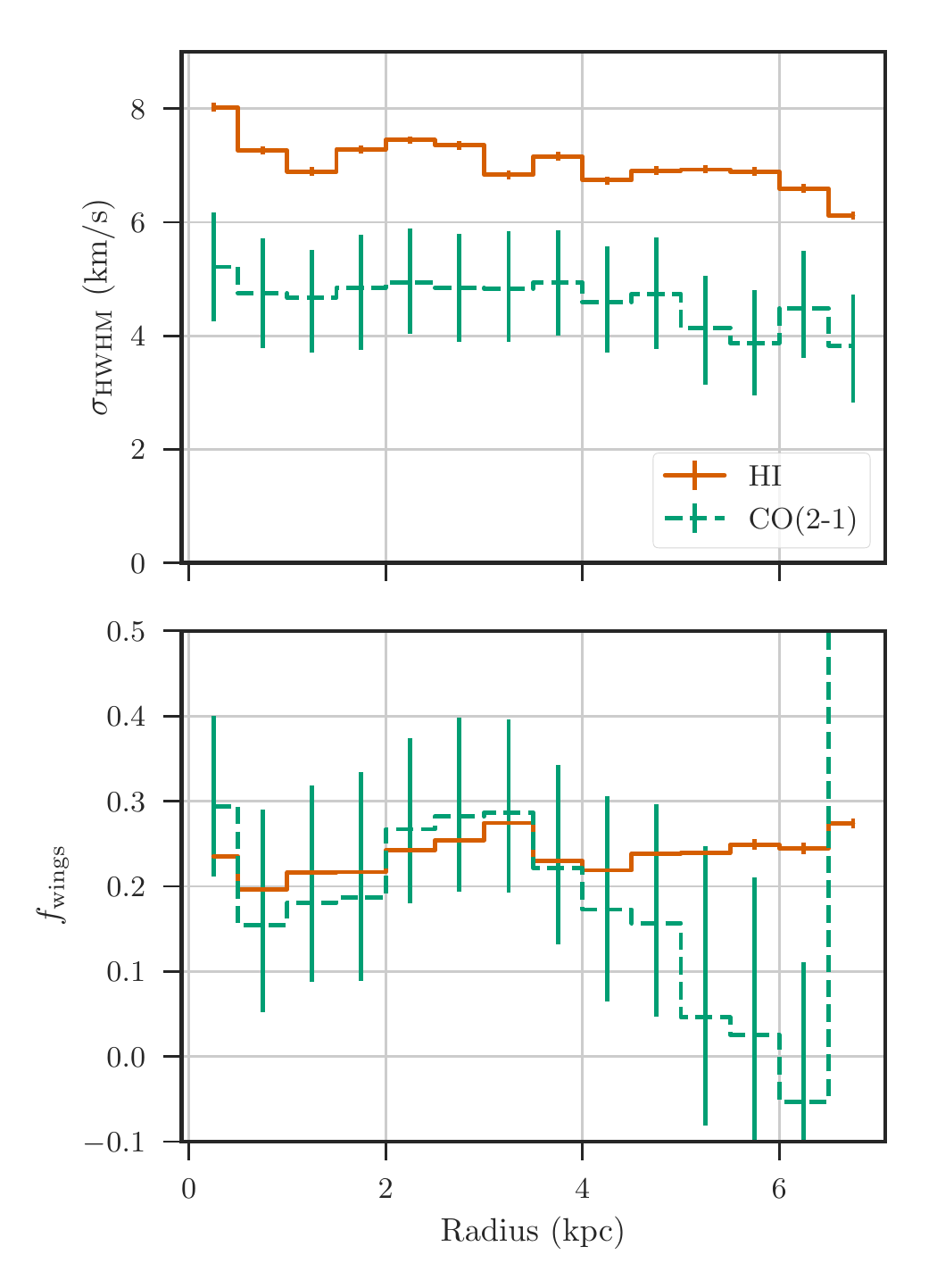}
\caption{\label{fig:radial_prof_widths} Stacked profile line widths ($\sigma_{\rm HWHM}$; top) and the fractional line wing excess ($f_{\rm wings}$; bottom) measured in $500$~pc radial bins. The widths are based on the HWHM approach from \citet{Stilp2013ApJ...765..136S}, and the errors are from a bootstrap approach described in the text.  The widths from both tracers show a shallow radial decline and have a consistent line width ratio of ${\sim}0.7$.}
\end{figure}

Figure~\ref{fig:radial_prof_widths} shows the line widths ($\sigma_{\rm HWHM}$) of the stacked profiles over the galactocentric radial bins (values provided in Table \ref{tab:hwhm_sigma_radial}).  We quantify the relation between galactocentric radius and the line widths by fitting a straight line. We exclude the innermost bins ($<1$~kpc) where beam smearing has a small contribution (Appendix \ref{app:line_broadening_from_beam_smearing}).  To account for the line width uncertainties, we resample the line widths in each bin from a Gaussian distribution with a standard deviation set by the uncertainty (Table \ref{tab:hwhm_sigma_radial}) in 1000 iterations.  We then estimate the slope and its uncertainty using the 15$^{\rm th}$, 50$^{\rm th}$, and 85$^{\rm th}$ percentiles from the distribution of 1000 fits.  We find that the \hi line widths decrease with galactocentric radius ($-0.14\pm0.01$ \kms kpc$^{-1}$), consistent with the stacking analysis in 100~pc bins by \citetalias{Koch2018MNRAS}. The decrease in the \co line widths with radius is insignificant at the 1-$\sigma$ level ($-0.16\pm0.16$~\kms kpc$^{-1}$).  We note, however, that cloud decomposition studies of the \cotwoone find a shallow line width decrease with galactocentric radius \citep{Gratier2012A&A...542A.108G,Braine2018A&A...612A..51B}.

Many nearby galaxies have a similar shallow radial decline in the \hi and \co line widths, outside of the galaxy centres \citep{Caldu2013AJ....146..150C,Mogotsi2016AJ....151...15M}.  Enhanced line widths are observed in galactic centres that result from a significant increase in the molecular gas surface density or the presence of a bar \citep[e.g.,][]{Sun2018ApJ...860..172S}, or due to beam smearing where the gradient of the rotation velocity is significant on the scale of the beam.  M33 is a lower mass spiral galaxy and lacks a strong bar, making it likely that the moderate line width increase within $R_{\rm gal}<0.5$ kpc is due to beam smearing (Appendix \ref{app:line_broadening_from_beam_smearing}).

A clear difference between our results and those by other studies is that the ratio between the CO and \hi line width is ${\sim}0.7$, differing from the ratio of $\sim1.0$ typically measured in other systems on $0.2\mbox{-}0.7$ kpc \citep{Combes1997A&A...326..554C,Caldu2013AJ....146..150C}. Since the observation of comparable CO and \hi line widths is used as an indicator of a thick molecular gas disk, we discuss this topic in detail in \S\ref{sec:no_thick_molecular_disk_in_m33_}.

Most of the shape parameters from the HWHM model do not show significant trends with galactocentric radius for the \cotwoone spectra and are insignificant at the $1\mbox{-}\sigma$ level.  The line wing excess ($f_{\rm wings}$) is significant for radial bins within $R_{\rm gal} < 5$ kpc.  At larger radii, the line wings become less prominent, though the \co detection fraction also sharply decreases at these radii and systematic effects---for example, from baseline fitting---strongly affect the stacked profile shapes beyond the HWHM.  This leads to the negative line wing excess at $6<R_{\rm gal}<6.5$ kpc and the large excess in the $6.5<R_{\rm gal}<7$ kpc bin.  The stacked profiles in Figure 12 of \citet{Druard2014A&A...567A.118D} also clearly suffer from these effects.

In the previous section, we estimated that error beam pick-up can account for up to $45\%$ of $f_{\rm wings}$ in the stacked profiles for $R_{\rm gal}<7$~kpc. Due to galactic rotation, the error-beam pick up will be asymmetric between the halves of the galaxy.  To test for this asymmetry, we also stack spectra in galactocentric radial bins separated into the northern and southern halves.  Though with significant uncertainties, the asymmetry of the northern half stacked profiles is consistently more negative than the southern half stacked profiles, which have asymmetries that are either positive or near zero.  This discrepancy between the halves shows that error beam pick-up accounts for some of the line wing excess.

The other model parameters are consistent between the northern and southern halves of M33.

\subsubsection{Stacked profiles at coarser resolution}
\label{subsub:stacked_profile_versus_spatial_scale}

We further investigate how stacked profile properties change with spatial resolution by repeating our analysis on data smoothed to a resolution of $160$~pc ($38\arcsec$) and $380$~pc ($95\arcsec$).  This allows for a more direct comparison to studies of stacked profiles on larger physical scales \citep{Caldu2013AJ....146..150C}.  At each resolution, we recompute the \hi peak velocities and create stacked \hi and \co profiles at that resolution.  Table~\ref{tab:global_profiles} gives the HWHM model parameters for these lower-resolution stacked profiles.

The line widths of both \hi and \cotwoone increase at coarser spatial resolution.  Based on stacking over the entire galaxy within $R_{\rm gal}=7$ kpc, we find $\sigma_{\rm HI}=8.0\pm0.1$ \kms and $\sigma_{\rm CO}=5.9\pm0.9$ \kms at a resolution of 160~pc, and $\sigma_{\rm HI}=8.9\pm0.1$ \kms and $\sigma_{\rm CO}=7.2\pm0.9$ \kms at a resolution of 380~pc.  The \cotwoone line widths have a larger relative increase than the \hi ones, which results in increased ratios of $\sigma_{\rm CO} / \sigma_{\rm HI} = 0.70\pm0.18$, $0.74\pm0.15$ and $0.81\pm0.12$ at scales of 80, 160 and 380~pc data, respectively.  However, the large uncertainties on the line width ratios makes this increase insignificant at the 1-$\sigma$ level. Using \co observations with higher spectral resolution will decrease these uncertainties and can determine whether this trend is significant.

There are two sources of line broadening that affect $\sigma_{\rm HWHM}$ as the resolution becomes coarser: (i) the dispersion between the \hi and \cotwoone peak velocities, and (ii) beam smearing.  Line broadening from the former source is due to aligning the \cotwoone spectra by the \hi peak velocity.  At a resolution of 80~pc, we estimate the standard deviation of the peak velocity difference to be $2.7$~\kms, as described in \S\ref{sub:peak_velocity_relation}.  Using the same procedure at the coarser resolutions, we find 1-$\sigma$ standard deviations of $3.1$ and $3.3$~\kms at a resolution of 160 and 380~pc, respectively.  The increase in the peak velocity difference moderately increases with resolution, but cannot account for the increase in the line widths at coarser resolution.

\begin{figure}
\includegraphics[width=0.5\textwidth]{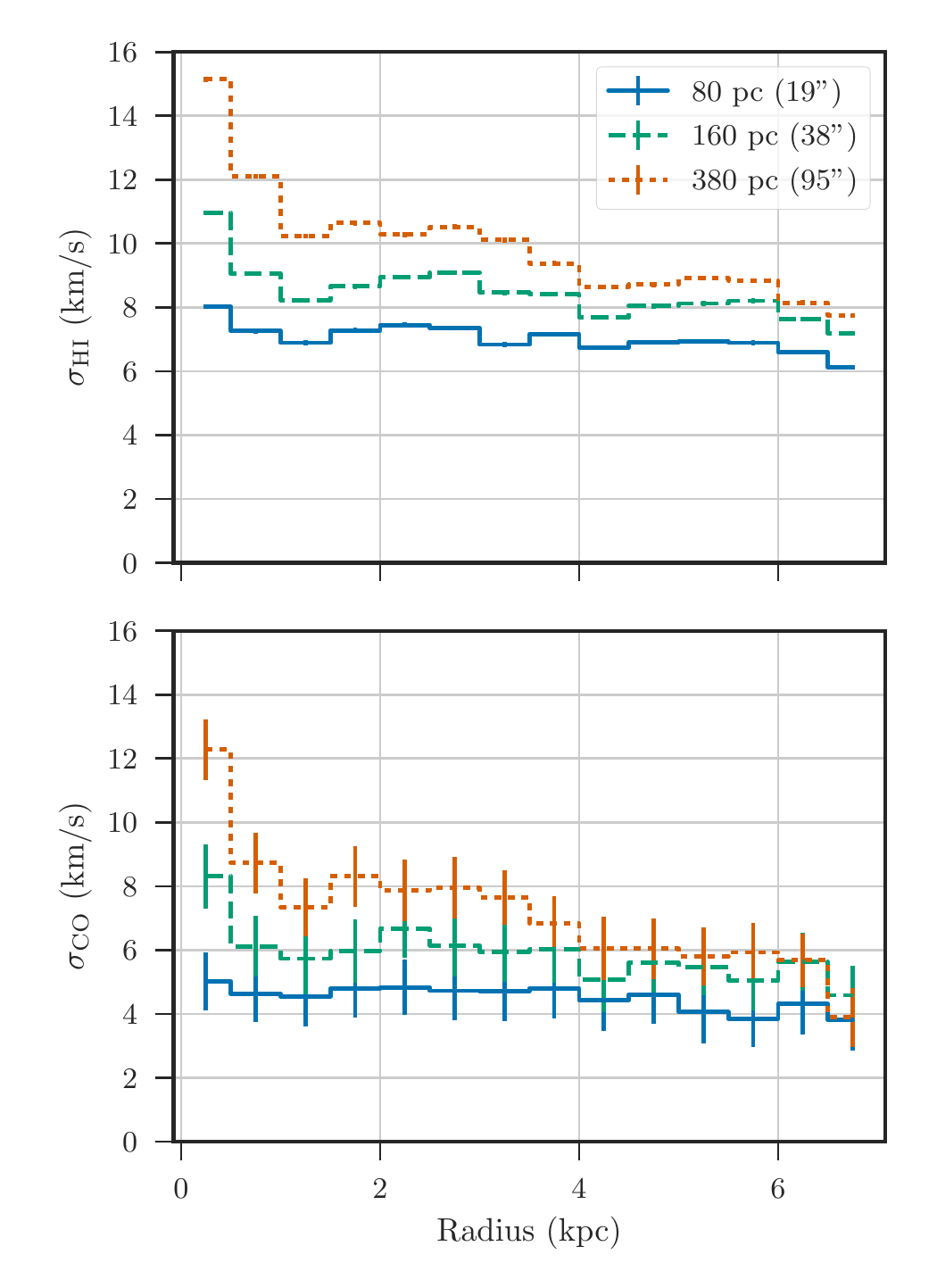}
\caption{\label{fig:radial_prof_widths_rescompare} Stacked profile line widths ($\sigma_{\rm HWHM}$) measured in $500$~pc radial bins at three different spatial resolutions for \hi (top) and \cotwoone (bottom). Line widths measured at coarser resolution have a steeper radial gradient due to beam smearing (Appendix \ref{app:line_broadening_from_beam_smearing}).}
\end{figure}

To address increased line broadening from beam smearing at coarser resolution, we repeat the line stacking in $500$~pc radial bins at each resolution.  Figure \ref{fig:radial_prof_widths_rescompare} shows the stacked line widths ($\sigma_{\rm HWHM}$) at the three spatial resolutions.  As the resolution becomes coarser, there is a steeper radial gradient in the line widths of both \hi and \co, particularly for $R_{\rm gal}<1$~kpc.  This radial trend qualitatively matches our estimate of beam smearing from Appendix \ref{app:line_broadening_from_beam_smearing}.
We determine how much of the line width increase with resolution is due to beam smearing with the area-weighted line broadening estimates calculated in Appendix \ref{app:line_broadening_from_beam_smearing}. For resolutions of 80 and 160~pc, the broadening from beam smearing is similar, with estimates of $2.0^{+2.1}_{-1.8}$~\kms and $1.5^{+1.7}_{-0.8}$~\kms, respectively.  The similar levels of beam smearing at 80 and 160~pc imply that the increase in the line width with resolution is not due to beam smearing.

At a resolution of 380~pc, beam smearing contributes significantly to the stacked line widths.  The area-weighted line broadening from beam smearing is $2.8^{+1.0}_{-1.0}$~\kms.  Treating the stacked profiles as Gaussian within the HWHM, we assume the line broadening can be subtracted in quadrature from the line width.  Applying this correction gives line widths of $8.4\pm0.9$~\kms for \hi and $6.6\pm3.6$~\kms for \cotwoone. The \cotwoone line width does not constrain whether the increased line widths are from beam smearing due to the uncertainty from the channel width.  However, the $0.9$~\kms increase in the \hi line width between the 160 and 380~pc data is much larger than the uncertainty and can entirely be explained by beam smearing.

On scales of 380~pc and larger, beam smearing becomes the dominant contribution to the line width.  With increasing scale, the stacked \hi and \co profiles will approach a common width since the stacking is performed with respect to the \hi and will be set by the rotation curve.  On smaller scales, systematics in the stacking procedure and beam smearing cannot account for the measured increase in the line widths.

We find that the \co line widths remain smaller than the \hi on scales up to $380$~pc, which is within the range where the stacking study by \citet{Caldu2013AJ....146..150C} find equivalent \co and \hi line widths.  We discuss this difference in the ratio of the line widths in M33 to other nearby galaxies in \S\ref{sec:no_thick_molecular_disk_in_m33_}.


\subsection{\hibold-CO Line of Sight Comparison} 
\label{sub:hi_line_widths_associated_with_co}

Stacked profiles provide a high S/N spectrum whose properties trace the average of the ensemble of stacked spectra. However, stacking removes information about the spatial variation of individual spectra and distributions of their properties.  By fitting individual lines-of-sight, the distributions of line shape parameters that lead to the shape of the stacked spectrum can be recovered.

Describing the \hi line profiles with an analytic model is difficult.  As we demonstrate in \citetalias{Koch2018MNRAS}, the typical \hi line profile in M33 is non-Gaussian due to multiple Gaussian components and asymmetric line wings.  The weak relations between the \co and \hi integrated intensities found by \citet{Wong2009ApJ...696..370W} in the LMC suggest that much of the \hi emission may be unrelated to the \co.  With this in mind, and since the \co spectra can typically be modelled with a single Gaussian at this resolution and sensitivity, we use the location of \co emission as a guide to decompose the \hi spectra and model the component most likely related to the \co.  This is an approximate method; a proper treatment of modeling individual \hi spectra requires a robust Gaussian decomposition \citep{Lindner2015AJ....149..138L,Henshaw:2016el}, which is beyond the scope of this paper.

\subsubsection{Fitting individual spectra}
\label{subsub:fitting_individual_spectra}

We relate the \hi and \co by using a limited decomposition of the \hi based on the spatial and spectral location of \cotwoone emission.  Towards lines-of-sight with \co detections, we determine the parameters of the single Gaussian component that is most closely related to the \co emission.

There are $19{,}796$ spectra where \cotwoone emission is detected above 3-$\sigma$ in three consecutive channels and is within $R_{\rm gal}<7$~kpc.  We model these spectra with the following steps:

\begin{enumerate}
  \item We fit a single Gaussian to the \cotwoone spectrum, accounting for broadening due to the channel width and the spectral response function using forward-modelling, which is described in Appendix \ref{app:forward_model}.  Forward-modelling accounts for line broadening due to the channel width and the spectral response function of the \cotwoone data (\S\ref{sub:iram_30_m}).
  \item The peak velocity of the \cotwoone fit defines the centre of a search window to find the nearest \hi peak.  The window is set to a width of three times the FWHM of the \cotwoone fitted width.
  \item  Since narrow extragalactic \hi spectra have widths $>2$~\kms \citep{Warren2012ApJ...757...84W}, we first smooth the \hi spectrum with a $2$~\kms box-car kernel.  Within the search window, we identify the closest \hi peak to the \cotwoone peak velocity\footnote{We did not find evidence for flattened \hi spectra in \citetalias{Koch2018MNRAS} on $80$~pc scales due to self-absorption so we do not search for absorption features aligned with the \co.}.
  \item Using the peak temperature of the identified \hi peak, we search for the HWHM points around the peak to define the \hi fit region.
  \item We fit a Gaussian to the un-smoothed \hi spectrum within the HWHM points of the identified peak.
\end{enumerate}

This approach assumes that \hi spectra are comprised of a small number of Gaussian components with well-defined peaks.  Since these restrictions are severely limiting, we define a number of checks to remove spectra that do not satisfy the criteria.  A line-of-sight is included in the sample if it meets the following criteria:

\begin{enumerate}
   \item The \hi peak is within the \cotwoone search window, defined above.  Based on visual inspection, \hi spectra that contain velocity-blended Gaussian components near the \cotwoone peak will not satisfy this criterion, and our naive treatment will fail to identify a single \hi component.
   \item The \co line width is larger than one channel width ($2.6$~\kms).
   \item Ten faint ($T_{\rm HI, peak} < 15$~K) \hi spectra have a fitted peak associated with noise.  The resulting \hi fitted profiles have narrow widths ($\sigma < 3$~\kms) and are removed from the sample.
   \item The fitted \hi peak velocity falls within the HWHM region or its Gaussian width is smaller than $12$~\kms.  This step removes \hi spectra with velocity-blended Gaussian components that significantly widen the Gaussian width and are not treated correctly with this method. We set the width threshold based on visual inspection.
   \item The \cotwoone line width is less than $8$~\kms.  A small fraction of \cotwoone spectra have multiple velocity components, and their fits to a single Gaussian all yield widths larger than $8$~\kms.
 \end{enumerate}

These restrictions yield a clean sample of $15{,}153$ spectra that we analyze here, $76\%$ of the eligible spectra.  Table \ref{tab:los_fitting} gives the properties of the line width distributions.  The uncertainties from each fit are from the covariance matrix of the least-squares fit.  We further validate the use of single Gaussian fits in Appendix~\ref{app:validating_the_gaussian_decomposition} and show examples of the fitting procedure.

\begin{table}
\caption{\label{tab:los_fitting} Mean line widths from the line-of-sight spectral fitting at different resolutions.  The uncertainties correspond to the 15$^{\rm th}$ and 85$^{\rm th}$ percentiles, respectively.}
\centering
\input{tables/los_scale_linewidth.tex}
\end{table}

\subsubsection{Relations between fitted parameters}
\label{subsub:relations_between_fitted_parameters}

We now examine the distributions of fitted \hi and \co line parameters to identify which parameters are related.

The fitted peak velocities are strongly correlated (Kendall-Tau correlation coefficient of $0.97$), consistent with the peak velocity over the LOS shown in Figure~\ref{fig:co_peak_vel_offset}.  The agreement is improved, however, since the outliers ($>10$~\kms) in Figure~\ref{fig:co_peak_vel_offset} are removed by only fitting the closest \hi component rather than the brightest one. Comparing the fitted peak velocities, the largest velocity difference between the \hi and \co is $9.5$~\kms.  The standard deviation between the fitted peak velocities is $2.0$~\kms, narrower than the $2.7$~\kms standard deviation from the line-of-sight peak velocity distribution from  \S\ref{sub:peak_velocity_relation}.

We also find that the peak \hi temperatures where \cotwoone is detected increase from $R_{\rm gal}<2$~kpc to $R_{\rm gal}>2$~kpc.  The brightest peak \hi temperatures ($>80$~K) are primarily found in the spiral arms, or spiral arm fragments, at $R_{\rm gal}>2$~kpc.  The lack of spiral structure in the inner $2$~kpc may lead to the lack of peak temperatures over ${\sim}80$~K.  These results imply that the \hi and \co peak temperatures are not strongly correlated, consistent with the small correlation coefficient of $0.1$ we find using the Kendall-Tau test.  The weak correlation in peak temperatures is consistent with \citet{Wong2009ApJ...696..370W}, who find that \hi peak temperature is poorly correlated to \co detections in the LMC.

The peak \co temperature has a negative correlation with \co\ line width, as would be expected for a Gaussian profile with a fixed integrated intensity. However, other studies using these \cotwoone data do not recover this negative correlation. \citet{Gratier2012A&A...542A.108G} find a positive, though weak, correlation between the peak \co temperature and the line width from a cloud decomposition analysis.  A similar correlation is found by \citet{Sun2018ApJ...860..172S}, who estimate the line widths with the equivalent Gaussian width determined from the peak temperature and integrated intensity of a spectrum \citep{Heyer2001ApJ...551..852H,Leroy2016ApJ...831...16L}.  The discrepancy between our results and these other works is due to requiring three consecutive channels $3\mbox{-}\sigma$ above the rms noise.  This biases our LOS sample, leading to incomplete distributions in the peak temperature and integrated intensity.

\begin{figure*}
\includegraphics[width=0.9\textwidth]{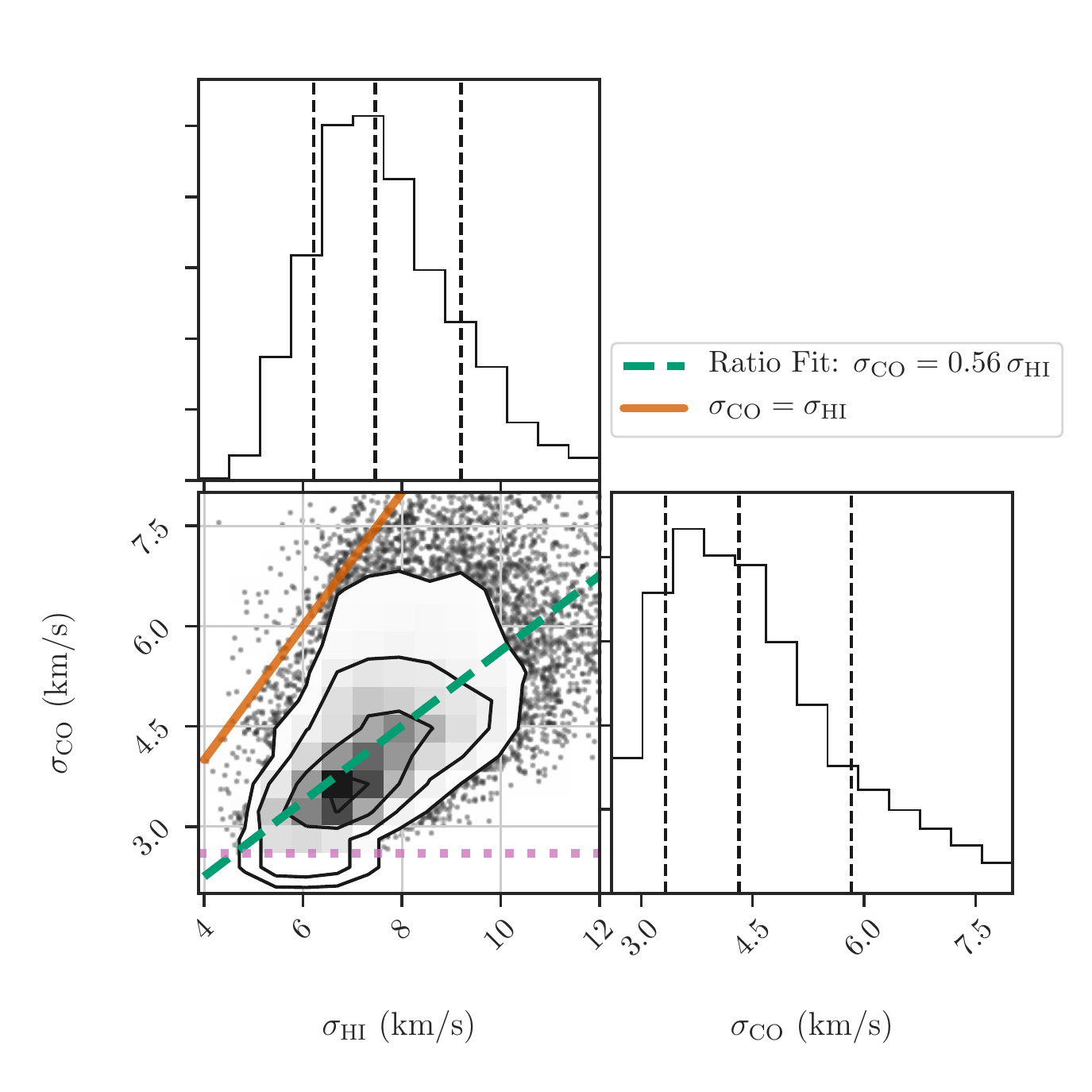}
\caption{\label{fig:component_linewidth_relation} Gaussian widths of individual \hi and \cotwoone profiles. The one-dimensional histograms show the distributions of $\sigma_{\rm HI}$ (top left) and $\sigma_{\rm CO}$ (bottom right) with vertical dashed lines indicating the 15$^{\rm th}$, 50$^{\rm th}$ and 85$^{\rm th}$ percentiles, respectively. The joint distribution is shown in the bottom left panel.  Contours show the area containing data within the 1- to 4-$\sigma$ limits of the distribution and black points show outliers beyond 4-$\sigma$, as described in Figure \ref{fig:co_peak_vel_offset}.  The orange solid line is the line of equality.  The horizontal dotted line indicates the \cotwoone channel width of $2.6$~\kms; no samples are included below this width. The green dashed line shows the fitted ratio of $0.56\pm0.01$. We note that our definition of a `clean' component sample restricts \hi line widths to be less than 12~\kms, and \co\ line widths must be less than 8~\kms. Typical uncertainties are $0.6$ and $0.2$~\kms for the \co and \hi, respectively.  A clear relation exists between the \hi and \co line widths with intrinsic scatter.}
\end{figure*}

Figure~\ref{fig:component_linewidth_relation} shows that there is a clear relation between $\sigma_{\rm CO}$ and $\sigma_{\rm HI}$.  Though there is significant dispersion in the relation, there is an increasing trend between the line widths of \cotwoone and \hi.  We find median line widths of $4.3$ and $7.4$~\kms for \co and \hi, respectively, on $80$~pc scales.  The \co line width distribution is near-Gaussian with a skew to large line widths, with 15$^{\rm th}$ and 85$^{\rm th}$ percentiles of $3.3$ and $5.8$~\kms.  The \hi distribution is more skewed to larger line widths compared to the \co distribution, and has 15$^{\rm th}$ and 85$^{\rm th}$ percentile values of $6.2$ and $9.2$~\kms.  The variations in $\sigma_{\rm HI}$ and $\sigma_{\rm CO}$ are larger than the typical uncertainties of $0.2$ and $0.6$~\kms, respectively.

We highlight the importance of restricting where the \hi is fit to in Appendix \ref{appsub:the_effect_of_hi_masking_on_fitting}, where we show that fitting the whole \hi spectrum leads to significant scatter in the \hi line widths that severely affects the relationships between the line widths we find here.

\citet{CalduPrimo2016AJ....151...34C} perform a similar restricted analysis of single Gaussian fitting to \co\ spectra of M31 at a deprojected resolution of $80\,{\rm pc} \times 380\,{\rm pc}$.  From their combined interferometric and single-dish data, they find typical \co\ line widths of $4.3\pm1.3$ \kms, consistent with the range we find.

We characterize the relationship between line widths by fitting for the line width ratio using a Bayesian error-in-variables approach \citep[see Section 8 of][]{Hogg2010arXiv1008.4686H}.  The goal of this approach is to fully reproduce the data in the model by incorporating (i) the line width uncertainties into the model and (ii) a parameter for additional scatter perpendicular to the line in excess of the uncertainties\footnote{This parameter tends to $0$ when no additional scatter is required to model the data.}.  We find that $\sigma_{\rm CO} = (0.56\pm0.01)\ \sigma_{\rm HI}$, which is shown in Figure~\ref{fig:component_linewidth_relation} as the the green dashed line.  The scatter parameter in the model is fit to be $0.52\pm0.02$~\kms, demonstrating that the scatter in the line width distributions exceeds the uncertainties.  This additional scatter represents real variations in the line width distributions.

We next examine whether changes in the line width with galactocentric radius can lead to additional scatter in the line width distributions.  Similar to the stacking analysis, we fit the line width relation within $500$~pc galactocentric bins out to a radius of $7$~kpc and find no variations in the average widths of the component with galactocentric radius, consistent with the shallow radial decrease from the stacked profile analysis (\S\ref{sub:total_profiles}).  By examining these possible sources of scatter in the line width distributions, we find that none of the sources can fully account for the scatter and that there must be additional variations not accounted for by the relationships of the fitted parameters.  We discuss the source of the scatter further in \S\ref{subsub:regional_variations_in_the_line_width_relation}.

\subsubsection{The Line Width Ratio at Coarser Spatial Resolution} 
\label{subsub:the_line_width_relation_at_lower_spatial_resolution}

Similar to the stacked profile analysis (\S\ref{subsub:stacked_profile_versus_spatial_scale}), we repeat the line-of-sight analysis when the data are smoothed to 160 and 380~pc.  The same fitting procedure is applied, with similar rejection criteria for poor fits.  However, we found that the line widths of valid fits to the data when smoothed to 380~pc can exceed the imposed cut-off values of $12$ and $8$ \kms for \hi and \cotwoone due to additional beam smearing on these scales (Appendix \ref{app:line_broadening_from_beam_smearing}).  Based on visual inspection, we increase these cut-off values to $17$~\kms and $12$~\kms.

Table \ref{tab:los_fitting} shows the line width distributions at these scales.  As we found with the stacked profile widths, the line widths increase at coarser resolution. The line width remains strongly correlated on these scales.

We fit for the line width ratios of the low-resolution samples and find values of $0.57\pm0.01$ and $0.63\pm0.01$ for the 160 and 380~pc resolutions, respectively.  The fitted ratios indicate that the line width ratio is relatively constant with increasing spatial resolution when analyzed on a line-of-sight basis.  We compare these line width ratios to those from the stacking analysis in \S\ref{sub:comparing_stacking_and_los}.

The line width ratios we find are moderately smaller than the line-of-sight analysis by \citet{Mogotsi2016AJ....151...15M} for a sample of nearby galaxies.  Fitting single Gaussians to \hi and \cotwoone spectra, they find a mean ratio of $0.7\pm0.2$ on spatial scales ranging from $200\mbox{--}700$~pc.  In contrast, the line width ratio we find is significantly steeper than extragalactic studies at higher physical resolution.  In the LMC, \citet{Fukui2009ApJ...705..144F} fit Gaussian profiles to both tracers where the \co\ peaks in a GMC and find a much shallower slope of $0.23$ at a resolution of $40$~pc.


\subsubsection{Regional Variations in the \hi-\co Line Widths } 
\label{subsub:regional_variations_in_the_line_width_relation}

\begin{figure}
\includegraphics[width=0.5\textwidth]{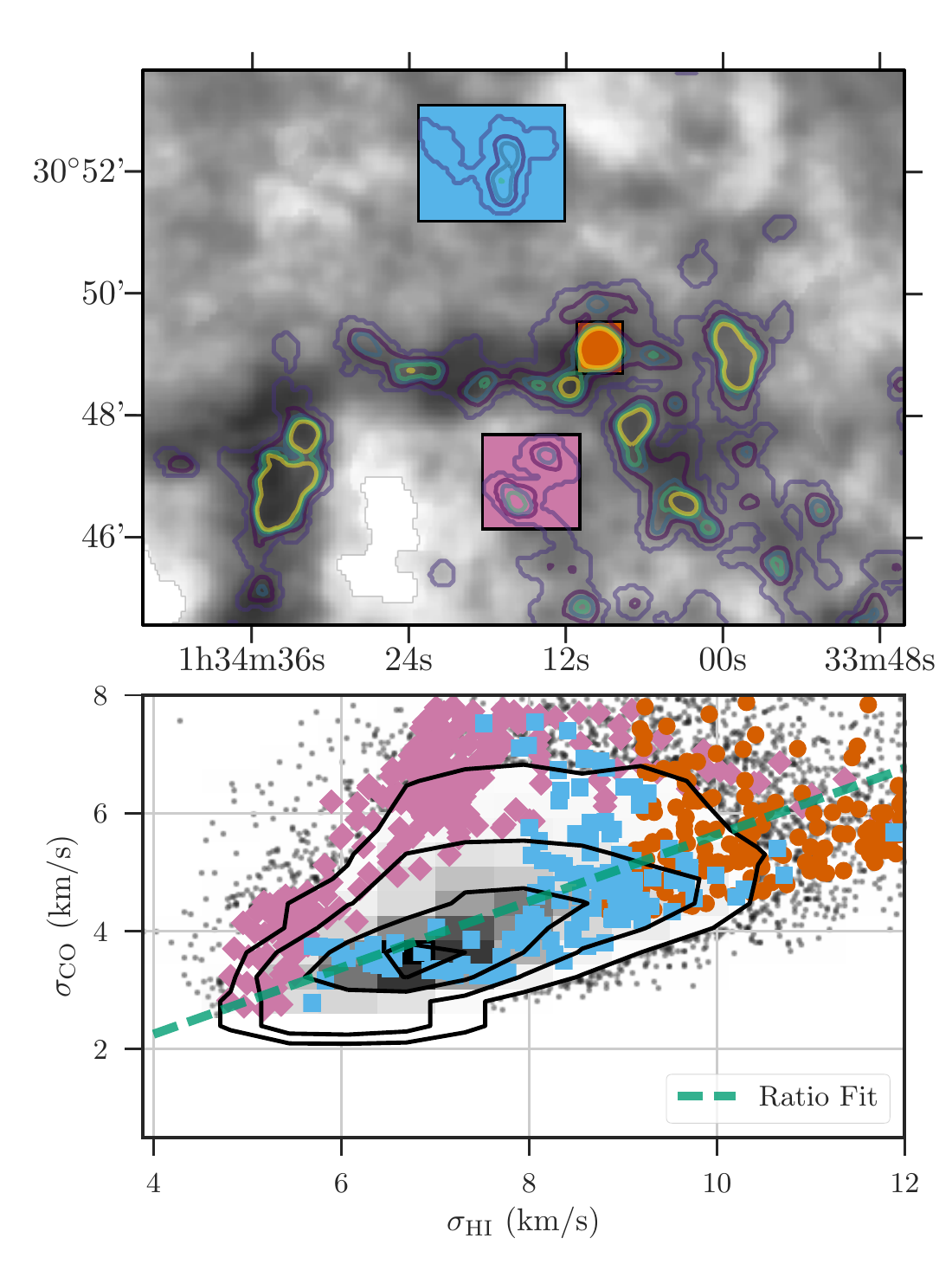}
\caption{\label{fig:linewidth_relation_regions} Top: The region around the Northern arm is shown, where the grayscale is the \hi column density and contours are the \cotwoone column densities with same levels shown in Figure~\ref{fig:hi_co_coldens}.  The coloured boxes indicate the line widths highlighted below. Bottom: The \hi-\cotwoone line width relation from Figure~\ref{fig:component_linewidth_relation} with line widths highlighted according to their region. The typical uncertainty of the fitted line widths is $0.2$~\kms for \hi and $0.6$~\kms  for \co.  The \hi and \co line widths remain correlated within individual regions, but their position in the line width plane varies with displacements larger than their uncertainties.  This suggests the line width relation is sensitive to environmental properties or the evolutionary state of GMCs.}
\end{figure}

To further investigate the observed correlation between \hi and \cotwoone line widths and the source of the scatter in this relationship, we highlight the positions of the line widths from three regions in Figure~\ref{fig:linewidth_relation_regions}.  These regions each have peak \co temperatures above the $75^{\rm th}$ percentile, and so the observed scatter is not driven by the correlation between peak \co temperature and line width.  By examining many regions on $\sim100$~pc scales, including the three examples shown, we find that the line widths remain correlated on these scales, but the slope and offset of the line widths varies substantially.  These regional variations are the source of the additional scatter required when fitting the \hi-\co line width relation (\S\ref{subsub:relations_between_fitted_parameters}).

By averaging over these local variations---over the full sample, in radial bins, and at different spatial resolutions---we consistently recover similar line width ratios.  The lack of radial variation indicates that $500$~pc radial bins provide a large enough sample to reproduce the scatter measured over the whole disk.  If these regional variations arise from individual GMCs, the \hi-\co line widths may indicate changes in the local environment or the evolutionary state of the cloud.  If the latter is true, the lack of a radial trend is consistent with the radial distribution of cloud evolution types from \citet{Corbelli2017A&A...601A.146C}, which are well-mixed in the inner $6$~kpc \citep[see also][]{Gratier2012A&A...542A.108G}.



\subsection{Spectral Properties from Stacking versus Individual Lines-of-sight}
\label{sub:comparing_stacking_and_los}

Previous studies of \hi and \co have found differing results between line stacking and fitting individual spectra.  While there are some discrepancies in the spectral properties we find, our results from these two methods are more similar than the results from other studies.
In this section, we explore the sources of discrepancy between the stacking and line-of-sight fit results and argue that most of the sources are systematic, due to the data or the analysis method.

The stacking analysis and line-of-sight fitting each have relative advantages and disadvantages. Stacked profiles provide an overall census of \hi and \cotwoone without conditioning on the spatial location of the emission.  However, variations in the centre and width of individual spectra---along with asymmetries and multiple components---will lead to larger line wings than a Gaussian profile of equivalent width.  This result is shown using a mixture model in \citetalias{Koch2018MNRAS}.

The line-of-sight (LOS) analysis retains spatial information, providing distributions of spectral properties that can be connected to different regions.  However, the simplistic decomposition of the \hi of this analysis requires the detection of \co along the line-of-sight, and so only provides an estimate of \hi properties where \co is detected. If \hi where \co is detected differs from the global population of \hi, the properties we find may not describe the typical \hi line properties.

We determine the source of discrepancies in our stacking and fitting results by creating stacked profiles from the fitted LOS sample and their Gaussian models.  There are five stacking tests we perform that are designed to control for variations in $\sigma_{\rm HWHM}$ or $f_{\rm wings}$. Table \ref{tab:los_stacking} gives the values for these parameters for each of the tests.  We described the purpose of each stacking test and their derived line profile properties below:
\begin{enumerate}
    \item {\it Stacked fitted model components aligned with the fitted CO mean velocity} --  The fitted Gaussian components, not the actual spectra, are stacked.  This removes all emission far from the line centre and will minimize $f_{\rm wings}$ in the stacked profiles.  Indeed, $f_{\rm wings}$ for \hi and \cotwoone are both significantly smaller than for the stacked profiles in \S\ref{sub:total_profiles} (Table \ref{tab:global_profiles}).  Stacking based on the fitted CO mean velocity will minimize the \co $\sigma_{\rm HWHM}$, while increasing the \hi $\sigma_{\rm HWHM}$ due to the scatter between the \hi and \co mean velocities.  The \co $\sigma_{\rm HWHM}$ is narrower than all of the other stacked profiles, including those from \S\ref{sub:total_profiles}, and is consistent with the mean \co LOS fitted width of $4.3^{+1.5}_{-1.0}$~\kms (Table \ref{tab:los_fitting}).
    \item {\it Stacked fitted model components aligned with the fitted HI mean velocity} --  This stacking test is identical to (i), except the fitted \hi mean velocities are used to align the spectra.  Aligning the spectra with the \hi mean velocity will decrease the \hi $\sigma_{\rm HWHM}$ and increase the \cotwoone $\sigma_{\rm HWHM}$, consistent with the measured properties.
    \item {\it Stacked spectra in the LOS sample aligned with the fitted HI mean velocity} -- The spectra in the LOS sample, rather than the model components used in (i) and (ii), are stacked aligned with the fitted \hi mean velocities.  The \co spectra in the sample are required to be well-modelled by a single Gaussian component, but there is a modest increase in $f_{\rm wings}$ to $0.07$., larger than in tests (i) and (ii)  The \hi spectra, however, have significant line structure that is not modelled for, leading to a vast increase in $f_{\rm wings}$ to $0.19$.  The line widths of the \hi and \co stacked profiles are the same within uncertainty.
    \item {\it Stacked spectra in the LOS sample aligned with the \hi peak velocity} -- The spectra used in (iii) are now aligned with the \hi peak velocities from \S\ref{sub:peak_velocity_relation}.  These stacked profiles are equivalent to the precedure used in \S\ref{sub:total_profiles} using only a sub-set of the spectra. This subset contains some of the outlier points in Figure \ref{fig:co_peak_vel_offset}, however $\sigma_{\rm HWHM}$ and $f_{\rm wings}$ do not significantly change from (iii). The outliers in the peak \hi and \co velocity difference distribution do not contribute significantly to $\sigma_{\rm HWHM}$ or $f_{\rm wings}$.
    \item {\it Stacked spectra in the entire LOS sample, including rejected fits, aligned with the \hi peak velocity} -- Finally, we create stacked profiles for the entire LOS sample considered in \S\ref{sub:hi_line_widths_associated_with_co}, including the LOS with rejected fits.  This test is equivalent to (iv) with a larger sample.  Relative to (iv), the \hi $\sigma_{\rm HWHM}$ and $f_{\rm wings}$ both moderately increase, as expected when including LOS potentially with multiple bright spectral components.  The \co stacked spectrum $\sigma_{\rm HWHM}$ marginally increases compared to (iv), however, $f_{\rm wings}$ increases by $33\%$ to $f_{\rm wings}=0.12$.  This increase is driven in part by the \co spectra with multiple components.
\end{enumerate}

\begin{table}
\caption{\label{tab:los_stacking} Stacked line width ($\sigma_{\rm HWHM}$) and line wing excess ($f_{\rm wings}$) from the spectra used in the line-of-sight analysis (\S\ref{subsub:fitting_individual_spectra}). The line widths do not strongly vary when changing the line centre definition or when the Gaussian model components are stacked rather than the actual spectrum. However,  $f_{\rm wings}$ is sensitive to whether the full spectra or the models are used. The \co $f_{\rm wings}$ is also more sensitive to the how the stacking is performed than the \hi.}
\centering
\input{tables/total_hwhm_los_stacking.tex}
\end{table}

From these tests, we can identify the source of the LOS and stacking discrepancies.

\subsubsection{Smaller \co LOS fitted line widths than from stacking}
\label{subsub:smaller_co_los_line_widths}

The larger \co stacked line widths are due to the scatter between the \hi and \co peak velocity (Figure \ref{fig:co_peak_vel_offset}).  This is demonstrated by comparing tests (i) and (ii), where the former is consistent with the median \co line width from the LOS fitting.  The larger \co line width from stacking will lead to an overestimate of the \hi-\co line width ratio.

\subsubsection{Larger \hi LOS fitted line widths compared with stacking}
\label{subsub:larger_hi_los_line_widths}

The stacked \hi line width towards LOS with \co detections (Table \ref{tab:los_stacking}) is consistently larger than the \hi stacked line width from all LOS (Table \ref{tab:global_profiles}).  There are two possible causes for this discrepancy. First, the \hi where \co is detected has larger line widths than the average from all \hi spectra. This source requires a physical difference in the atomic gas properties where molecular clouds are located, possibly related to the \hi cloud envelope \citep{Fukui2009ApJ...705..144F}.
Alternatively, the \hi components fitted here may be broadened by overlapping velocity components since our analysis does not account for this.  However, from visually examining the fits to the LOS sample, most \hi spectra would need to have highly overlapping components for the average \hi LOS line width to be broadened, and this does not seem likely for most spectra (Appendix \ref{appsub:examples_of_fitted_spectra}).  In order to definitely determine which of these sources leads to the larger \hi line widths, we require decomposing the \hi spectra without conditioning on the location of the \co emission. However, this analysis is beyond the scope of this paper.  We favour larger \hi line widths in molecular cloud envelopes as the source of this discrepancy.

\subsubsection{Sources of the line wing excess}
\label{subsub:sources_of_the_line_wing_excess}

These five tests provide restrictions on the source of the line wing excess in both tracers.  In the \hi, $f_{\rm wings}$ is only changed when the model components ((i) and (ii)) are stacked rather than the full \hi spectra ((iii)--(v)).  This result is consistent with the line structure and wings evident in individual \hi spectra, as explored in \citetalias{Koch2018MNRAS}. The scatter in the fitted velocities (\hi or \co) can account for $f_{\rm wings}\sim0.03$ in the stacked \hi profiles.

For the \co stacked profiles, there are variations in $f_{\rm wings}$ from multiple sources.  There are small contributions to $f_{\rm wings}$ from the scatter in the \hi fitted mean velocities (Test (ii); $f_{\rm wings}=0.03$) and the scatter between the \hi and \co peak velocities or fitted mean velocities (Tests (iii) \& (iv); $f_{\rm wings}=0.01$).  Multi-component \co spectra account for $f_{\rm wings} \lesssim 0.04\mbox{--}0.05$ from comparing Tests (iii) and (iv) to Test (v).  This estimate is an upper limit since we do not control for contributions from real line wings versus multiple spectral components.  Finally, comparing Tests (iii) and (iv) to Test (ii), excess line wings can directly account for $f_{\rm wings}=0.04\mbox{--}0.05$.

The different sources of line wing excess in the \co stacked profiles implies that there is marginal evidence for \co line wings.  As described above, stacking systematics and multi-component spectra can account for $f_{\rm wings} \lesssim 0.08\mbox{--}0.09$, roughly half of the line wing excess of $f_{\rm wings}= 0.21$ from the stacked profile towards all LOS (Table \ref{tab:global_profiles}).  The discrepancy between the total $f_{\rm wings}$ for the \co stacked profiles and the systematics is then $0.13$, though the large $f_{\rm wings}$ uncertainties can account for the remaining line wing excess.  Including the error beam contribution of up to $45\%$ of the line wing excess (\S\ref{sub:total_profiles}), we find that $\sim80\%$ of the \co line wing excess can be accounted for without requiring the presence of real \co line wings.  However, due to the estimated uncertainties, we cannot rule out their presence.

\section{A Marginal Thick Molecular Disk in M33} 
\label{sec:no_thick_molecular_disk_in_m33_}

Studies of \co in the Milky Way and nearby galaxies find evidence of two molecular components: a thin disk dominated by GMCs, and a thicker diffuse molecular disk.  Our results, however, suggest that M33 has a marginal thick molecular component, unlike those found in other more massive galaxies, based on (i) finding smaller \co line widths relative to the \hi and (ii) the marginal detection of excess \co line wings.  In this section, we compare our results to previous studies and address previous works arguing for a diffuse component on large-scales in M33.

Evidence for a diffuse molecular component has been demonstrated with extended emission in edge-on galaxies \citep[e.g., NGC~891;][]{GarciaBurillo1992A&A...266...21G}, separating $^{12}$CO emission associated with denser gas in the Milky Way \citep{RomanDuval2016ApJ...818..144R}, comparing the flux recovered in interferometric data to the total emission in single-dish observations \citep{Pety2013ApJ...779...43P,CalduPrimo2015AJ....149...76C,CalduPrimo2016AJ....151...34C}, and large \co line widths in nearby galaxies \citep{Combes1997A&A...326..554C,Caldu2013AJ....146..150C,CalduPrimo2015AJ....149...76C}.

In M33, a diffuse molecular component has been suggested based on the \co flux recovered in GMCs \citep{Wilson1990ApJ...363..435W}, comparing the \thco to $^{12}$CO spectral properties \citep{Wilson1994ApJ...432..148W}, and a non-zero \co power-spectrum index on kpc scales \citep{Combes2012A&A...539A..67C}.  \citet{Rosolowsky2007ApJ...661..830R} find that 90\% of the diffuse \co emission is located $<100$~pc to a GMC, and suggest the emission is due to a population of unresolved, low-mass molecular clouds \citep[see also][]{Roso2003ApJ...599..258R}.

\begin{figure}
\includegraphics[width=0.5\textwidth]{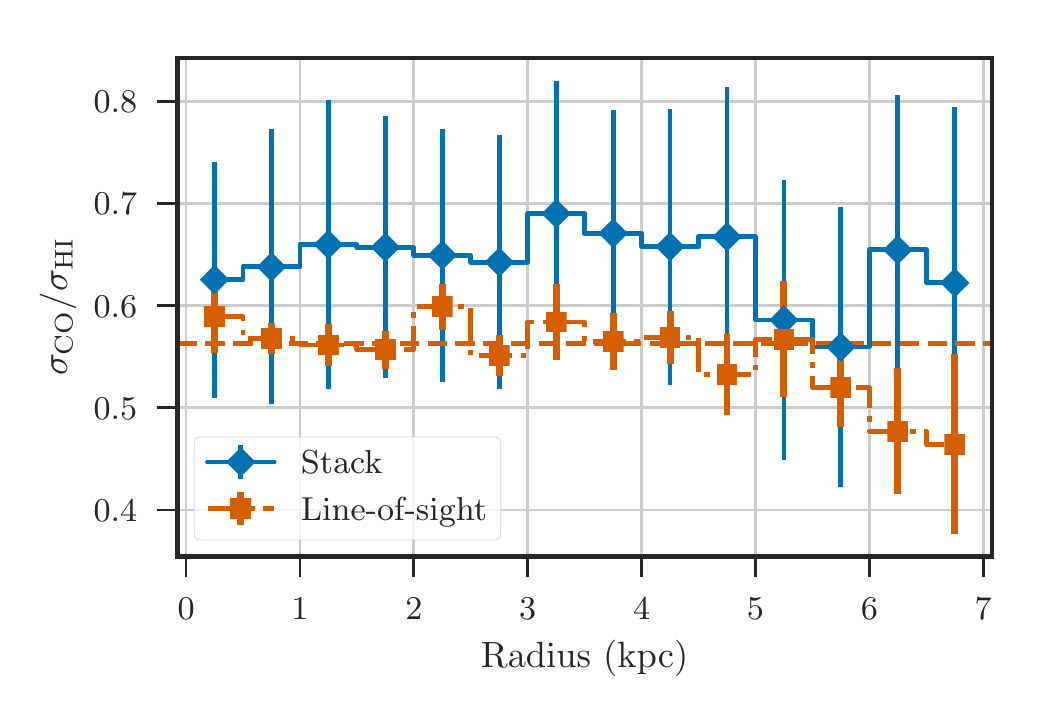}
\caption{\label{fig:ratio_vs_radius} Line width ratio from stacked profiles (blue solid diamonds) and the average from the line-of-sight fits (orange dot-dashed squares) versus galactic radius at $80$~pc ($20\arcsec$) resolution.  The 1-$\sigma$ uncertainties on the stacked widths are dominated by the \cotwoone channel width.  The errors on the line-of-sight fits are the standard deviation in the radial bin divided by the square root of the number of independent components.  The pink dashed line is the fitted ratio $0.56$ shown in Figure~\ref{fig:component_linewidth_relation}.  Both methods have line width ratios smaller than unity, unlike other (more massive) nearby galaxies \citep{Caldu2013AJ....146..150C}, suggesting M33 lacks a significant thick molecular disk.}
\end{figure}

Our stacking analysis, and the stacking analysis in \citet{Druard2014A&A...567A.118D}, shows that the \co line widths are consistently smaller than the \hi, unlike the line widths from most nearby galaxies on $\sim500$~pc scales \citep{Combes1997A&A...326..554C,Caldu2013AJ....146..150C}.  Figure~\ref{fig:ratio_vs_radius} summarizes our results by showing the \co-to-\hi line width ratios of the stacked line widths and the radially-binned averages from the line-of-sight analysis at $80$~pc resolution. The ratios from the stacked profiles are consistently $10\%$ larger than the average of the line-of-sight fits due to using the \hi peak velocities to align the \co spectra (\S\ref{sub:comparing_stacking_and_los}).  The line width ratio increases but remains less than unity when the data are smoothed to resolutions of $160$ and $380$~pc (\S\ref{subsub:stacked_profile_versus_spatial_scale}), scales comparable to some of the data in \citet{Caldu2013AJ....146..150C}.

Stacking analyses of \co by \citet{CalduPrimo2015AJ....149...76C} and \citet{CalduPrimo2016AJ....151...34C} find a wide Gaussian component to \co stacked profiles that arises only in single-dish observations.  The high-resolution ($80\times350$~pc) \co observations of M31 from \citet{CalduPrimo2016AJ....151...34C} constrain this wide component to scales of $\sim500$~pc and larger.  Coupled with the large \co line widths, these results suggest the wide Gaussian component is due to a thick molecular disk.  In our analysis, the wide Gaussian component would contribute to the line wing excess ($f_{\rm wings}$)\footnote{We stress that the flux in a wide Gaussian component will not equal $f_{\rm wings}$. However, an increase in the amplitude or width of the wide Gaussian component will be positively correlated with $f_{\rm wings}$.}.  In M33, we find a qualitatively similar line wing excess to \citeauthor{CalduPrimo2016AJ....151...34C}, however, up to $\sim80$\% of the excess is due to stacking systematics and error beam pick-up from the IRAM 30-m telescope \citep[\S\ref{subsub:sources_of_the_line_wing_excess};][]{Druard2014A&A...567A.118D}.  The remaining fraction of the \co line wing excess is small, and would correspond to a much smaller contribution from a wide Gaussian component compared to those found by \citet{CalduPrimo2015AJ....149...76C} and \citet{CalduPrimo2016AJ....151...34C}.

These results strongly suggest M33 has a marginal thick molecular disk and is instead more consistent with the findings from \citet{Rosolowsky2007ApJ...661..830R} where diffuse \co emission is clustered near GMCs and may be due to unresolved low-mass clouds \citep[flux recovery with spatial scale with these \co data is explored in][]{Sun2018ApJ...860..172S}.  There remains ambiguity about the diffuse molecular component in M33 from other analyses, and whether M33 is the only nearby galaxy with a marginal thick molecular disk.  We address these issues in the following sections.  First, we demonstrate that the large-scale \cotwoone power-spectrum identified in \citet{Combes2012A&A...539A..67C} can be explained by the exponential disk scale of the \co emission rather than a thick molecular disk.  We then note the similarity of the line width ratios found by \citet{Caldu2013AJ....146..150C} for NGC~2403 and our results.

\subsection{Comparison to a thick molecular disk implied by power-spectra}
\label{sub:comparison_to_a_thick_molecular_disk_implied_by_powerspectra}

Previous work by \citet{Combes2012A&A...539A..67C} found that the power-spectra of \hi and \co integrated intensity maps of M33 have shallow indices extending to kpc scales, with distinct breaks near $\sim120$~pc where the indices becomes steeper.  The non-zero slope on kpc scales suggests there is significant large-scale diffuse emission in M33 from both \hi and \co, in contrast with our findings for \co from the line widths.

To explain the non-zero index on large scales, we compare the large-scale distribution of emission in M33 for \co and \hi.  Bright \co\ emission is broadly confined to individual regions on scales comparable to the beam size (Figure~\ref{fig:hi_co_coldens}) and has a radial trend in the average surface density that is well-modelled by an exponential disk with a scale length of $\sim2$~kpc \citep{Gratier2010A&A...522A...3G,Druard2014A&A...567A.118D}.  This radial trend implies that the detection fraction per unit area of \co also depends on radius, providing additional power in the power-spectrum on $\sim$~kpc scales.  On the other hand, \hi emission is widespread throughout the disk and the surface density is approximately constant within the inner 7~kpc \citepalias{Koch2018MNRAS}.   The difference in the large-scale radial trends of \hi and \co will affect the large-scale ($\sim$~kpc) parts of the power-spectrum.

\begin{figure}
\includegraphics[width=0.5\textwidth]{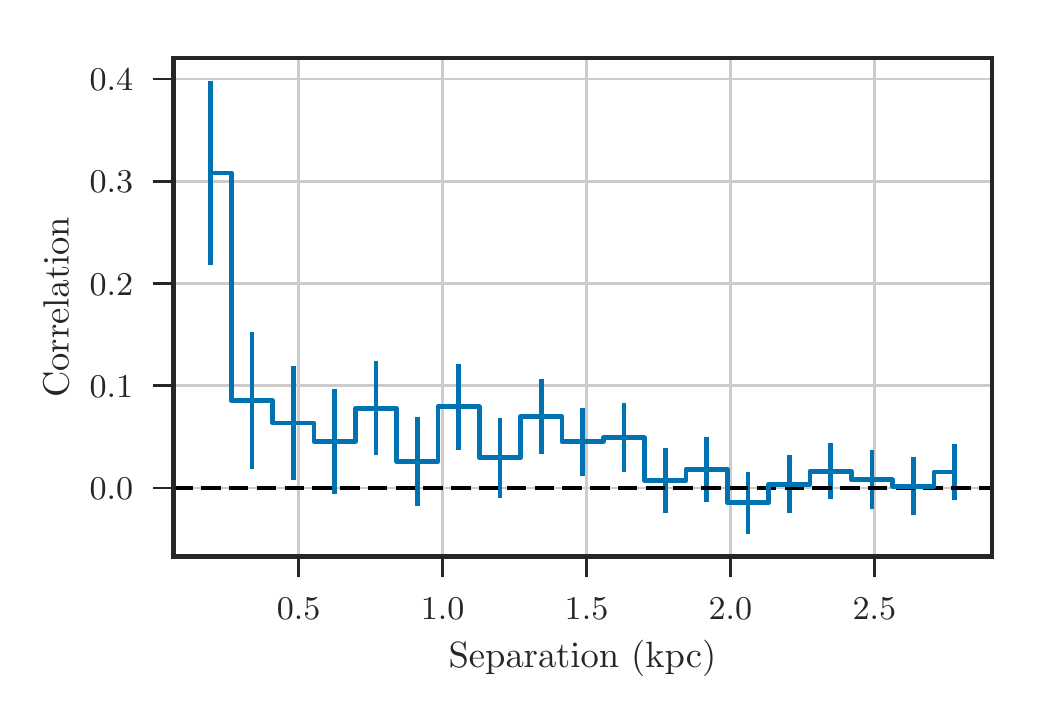}
\caption{\label{fig:2ptcorr} Two-point correlation function of the GMC positions from \citet{Corbelli2017A&A...601A.146C} measured in 150~pc bins.  The uncertainties are estimated from 2000 bootstrap iterations.  Most of the correlation occurs on $<150$~pc scales, but the structure of \co emission due to the disk gives non-zero correlations up to $\sim2$~kpc scales.}
\end{figure}

We demonstrate how the exponential \co disk affects the power-spectrum by calculating the two-point correlation function of the GMC positions from \citet{Corbelli2017A&A...601A.146C}.  By treating the \co emission as a set of point sources at the GMC centres, any large-scale correlations must result from spatial clustering, rather than extended \co emission. Figure \ref{fig:2ptcorr} shows that the two-point correlation function of the cloud positions has a non-zero correlation up to scales of $\sim2$~kpc, similar to the exponential \co disk scale.  This non-zero correlation on these scales will correspond to a non-zero \co\ power spectrum slope, thus demonstrating that the large-scale power-spectrum does not imply the presence of a thick molecular disk.

This result may also explain the large change in the \co power-spectrum index across the $\sim120$~pc break point found by \citet{Combes2012A&A...539A..67C}. Distinct breaks in the power-spectra, and other turbulent metrics, are useful probes of the disk scale height \citep{Elmegreen2001ApJ...548..749E,Padoan2001ApJ...555L..33P}.  The power-spectrum index is predicted to change by $+1$ on scales larger than the disk scale height since large-scale turbulent motions are confined to two dimensions in the disk \citep{Lazarian2000ApJ...537..720L}.   \citet{Combes2012A&A...539A..67C} find the \co power-spectrum index changes by $+2.2$ across the break, significantly larger than the expected change of $+1$.  For the \hi, the index change of $+0.8$ across the break is much closer to the expected change.

The similarity \citet{Combes2012A&A...539A..67C} find between the \co and \hi break points in the power-spectra also differs from the disk scale heights implied by the line widths we find.  For ratios $\sigma_{\rm CO}/\sigma_{\rm HI} < 1$, the disk scale height of \co should be smaller than the \hi; we can approximate the ratio of the disk scale heights from the line width ratio.  For $R_{\rm gal}<7$~kpc, the stellar surface density is larger than the total gas surface density in M33 \citep{Corbelli2014A&A...572A..23C}.  Measurements of the stellar velocity dispersion find $\sim20$~\kms in the inner disk \citep{Kormendy1993AJ....105.1793K,Gebhardt2001AJ....122.2469G,Corbelli2003MNRAS.342..199C}, suggesting the stellar disk scale height is larger than the \hi and \co disk scale heights.  If this result holds true for $R_{\rm gal}<7$~kpc, the ratio of the \co and \hi disk scale heights is the ratio of the line widths: $H_{\rm CO} / H_{\rm HI} \approx \sigma_{\rm CO} / \sigma_{\rm HI}$ \citep{Combes1997A&A...326..554C}, for line widths measured at the disk scale height.  Based on our analysis at $80$ and $160$~pc, the line width ratio is $\sim0.6$, suggesting that the \co scale height should be $\sim60\%$ of the \hi scale height\footnote{The line width ratio from the LOS analysis may be too small, due to the large \hi line widths (\S\ref{subsub:larger_hi_los_line_widths}), while the ratio from the stacking analysis is too large due to the \co stacked line width being larger than the LOS analysis (\S\ref{subsub:smaller_co_los_line_widths}). A ratio of $\sim0.6$ is in-between the ratios from the two methods.}.

The discrepancy with the scale heights we measure and the similar scale of the break points from \citet{Combes2012A&A...539A..67C} may be a limitation of the data resolution used in their analysis.  They use \hi and \co data at $12\arcsec$ ($\sim50$~pc) resolution \citep{Gratier2010A&A...522A...3G} and the constraints on the scale of the break are limited by the beam size, as shown in their Appendix B. Higher resolution observations ($\sim20$~pc) should determine whether there is a difference in the disk scale heights traced by \hi and \co.

\subsection{Similarities with the flocculent spiral NGC~2403}
\label{sub:similarities_with_2403}

From the sample of nearby galaxies studied in \citet{Caldu2013AJ....146..150C}, there are two galaxies that are also dominated by the atomic component throughout most of the disk: NGC~925 and NGC~2403 \citep[see also][]{Schruba2011AJ....142...37S}.  As these are the closest analogs to M33 in their sample, we compare the stacked profile analyses from \citet{Caldu2013AJ....146..150C} to our results.

The NGC~925 line width ratios are consistent with unity, however, the signal-to-noise in the \co\ map limit the analysis to a few radial bins at the galaxy centre.  The signal-to-noise of the NGC~2403 data is higher, allowing the analysis to be extended to larger radii ($70\%$ of the optical radius) providing a number of radial bins for comparison.  NGC~2403 is also the closest galaxy in their sample and the physical scale of the beam is ${\sim}200$~pc, a factor of about two coarser than the resolution of our M33 data.  Interestingly, the line width ratios, outside of the galactic centre (${\sim}0.1R_{25}$), are consistently smaller than unity, with an average of ${\sim}0.8$.  The increased line widths in the inner disk are likely affected by beam smearing.  With the same data, the line-of-sight analysis by \citet{Mogotsi2016AJ....151...15M} find a smaller line width ratio of ${\sim}0.7$.  Both of these line width ratios are comparable to our results in M33 on $160$~pc ($38\arcsec$) scales.  Our results are then consistent with the ratios for NGC~2403 from \citet{Caldu2013AJ....146..150C} and \citet{Mogotsi2016AJ....151...15M}, suggesting that galaxies with atomic-dominated neutral gas components have at most a moderate contribution from a diffuse molecular disk.



\section{Summary}
\label{sec:summary}

We explore the spectral relationship of the atomic and molecular medium in M33 on 80~pc scales by comparing new VLA \hi observations \citepalias{Koch2018MNRAS} with IRAM \mbox{30-m} \cotwoone data \citep{Gratier2010A&A...522A...3G,Druard2014A&A...567A.118D}.
We perform three analyses---the difference in the velocity at peak \hi and \cotwoone brightness, spectral stacking, and fits to individual spectra---to explore how the atomic and molecular ISM are related.  Each of these analyses demonstrates that the spectral properties between the \hi and \co are strongly correlated on $80$~pc scales.  We also show that relationship between the \hi and \co line widths, on $80$~pc scales, from individual spectral fits depend critically on identifying the \hi most likely associated with \co emission, rather than all \hi emission along the line-of-sight.

\begin{enumerate}

  \item The velocities of the \hi and \co\ peak temperatures are closely related.  The standard deviation in the differences of these velocities is $2.7$~\kms, slightly larger than the \co\ channel widths ($2.6$~\kms; Figure~\ref{fig:co_peak_vel_offset}).  Significant outliers in the velocity difference ($>10$~\kms) occur where the \hi spectrum has multiple components and the \co\ peak is not associated with the brightest \hi peak.  These outliers are removed when modelling only the \hi component associated with \co\ (\S\ref{sub:hi_line_widths_associated_with_co}).

  \item By stacking \hi and \co\ spectra aligned to the velocity of the peak \hi brightness, we find that the width of the \co\ stacked profile ($4.6\pm0.9$ \kms) is smaller than the \hi stacked profile ($6.6\pm0.1$ \kms) on 80~pc scales, unlike similar analyses of other (more massive) nearby galaxies that measure comparable line widths on 500~pc scales \citep[e.g.,][]{Caldu2013AJ....146..150C}.  The widths of the stacked profiles slowly decrease with galactocentric radius.

  \item By repeating the stacking analysis at lower spatial resolutions of 160~pc ($38\arcsec$) and 380~pc ($95\arcsec$), we find that the \cotwoone-to-\hi line width ratio remains constant within uncertainty.  We estimate how beam smearing contributes at each resolution and find that resolutions of 80 and 160~pc have a similar contribution to the line width from beam smearing. Beam smearing contributes more at a resolution of 380~pc and can explain the increased line widths relative to those at 160~pc.  However, the \co line width remains smaller than the \hi on all scales.

  \item We perform a spectral decomposition of \hi spectra limited to where \co is detected.  The \co spectra are fit by a single Gaussian, while the Gaussian fit to the \hi is limited to the closest peak in the \hi spectrum.  We carefully inspect and impose restrictions to remove spectra where this fitting approach is not valid.  The average \hi and \co\ line widths of the restricted sample are $7.4^{+1.7}_{-1.3}$ and $4.3^{+1.5}_{-1.0}$~\kms, where the uncertainties are the 15$^{\rm th}$ and 85$^{\rm th}$ percentiles of the distributions, respectively.

  \item The average \co line width from the line-of-sight fits are smaller than those from the stacking analysis.  This difference results from aligning the \co spectra to the \hi peak velocity, while there is scatter between the \hi and \co velocity at peak intensity (Figure~\ref{fig:co_peak_vel_offset}).  Recovering larger \co stacked line widths relative to those from individual spectra is a general result that will result whenever \co is aligned with respect to another tracer, such as \hi.  The amount of line broadening is set by the scatter between the line centres of the two tracers.  Thus, line stacking based on a different tracer will bias the line widths to larger values, but is ideal for recovering faint emission \citep{Schruba2011AJ....142...37S}.

  \item The average \hi line width from the line-of-sight fits ($7.4\pm1.5$~\kms) is {\it larger} than the stacked profile width ($6.6\pm0.1$~\kms).  The larger line widths are due to either multiple highly-blended Gaussian components that are not modelled correctly in our analysis, or that the \hi associated with \co emission tends to have larger line widths.  We favour the latter explanation since our line-of-sight analysis has strong restrictions to remove multi-component spectra (Appendix \ref{app:validating_the_gaussian_decomposition}); however, we do not fully decompose the \hi spectra and cannot rule out the former explanation.

  \item The line-of-sight fits highlight a strong correlation between \hi and \co\ line widths (Figure~\ref{fig:component_linewidth_relation}).  We fit for the line width ratio, accounting for errors in both measurements, and find $\sigma_{\rm CO} = (0.56\pm0.01) \ \sigma_{\rm HI}$, smaller than the ratios from the stacked profiles due to the smaller average \co line width and larger average \hi line width.  There is no trend between the line width ratio and galactocentric radius (Figure \ref{fig:ratio_vs_radius}).  When repeated at a lower spatial resolution, we find that the \hi and \co\ line widths are increased by the same factor, leading to the same line width ratio, within uncertainties.

  \item The scatter in the relation between the \hi and \co line widths is larger than the statistical errors and results from regional variations (Figure~\ref{fig:linewidth_relation_regions}). The line widths of \hi and \co remain correlated when measured in individual regions, but exhibit systematic offsets with respect to the median \hi and \co line widths.  These regional variations affect both the \hi and \co\ line widths and suggest that the local environment plays an important role in setting the line widths.

  \item We perform stacking tests with the fitted LOS components to constrain sources of the line wing excess. We find that the error beam pick-up from the IRAM \mbox{30-m} telescope \citep{Druard2014A&A...567A.118D} and systematics of the stacking procedure can account for $\sim80$\% of the line wing excess in the \cotwoone (\S\ref{subsub:sources_of_the_line_wing_excess}).  Combined with a \co-to-\hi line width ratio less than unity, this result implies that M33 has at most a marginal thick molecular disk.  We point out that previous analyses of NGC 2403 give similar results to ours \citep{Caldu2013AJ....146..150C,Mogotsi2016AJ....151...15M}, suggesting that galaxies where the atomic component dominates the cool ISM may lack a significant thick molecular disk.

\end{enumerate}

Scripts to reproduce the analysis are available at \url{https://github.com/Astroua/m33-hi-co-lwidths}\footnote{Code DOI: \url{https://doi.org/10.5281/zenodo.2563161}}.

\section*{Acknowledgments}

We thank the referee for their careful reading of the manuscript
and comments. EWK is supported by a Postgraduate Scholarship from the Natural Sciences and Engineering Research Council of Canada (NSERC). EWK and EWR are supported by a Discovery Grant from NSERC (RGPIN-2012-355247; RGPIN-2017-03987).  This research was enabled in part by support provided by WestGrid (\url{www.westgrid.ca}), Compute Canada (\url{www.computecanada.ca}), and CANFAR (\url{www.canfar.net}). The work of AKL is partially supported by the National Science Foundation under Grants No. 1615105, 1615109, and 1653300. The National Radio Astronomy Observatory and the Green Bank Observatory are facilities of the National Science Foundation operated under cooperative agreement by Associated Universities, Inc.

\textbf{Code Bibliography: }
CASA \citep[versions 4.4 \& 4.7;][]{casa_mcmullin2007} --- astropy \citep{astropy} --- radio-astro-tools (spectral-cube, radio-beam; \url{radio-astro-tools.github.io}) --- matplotlib \citep{mpl} ---  seaborn \citep{seaborn} --- corner \citep{corner} --- astroML \citep{astroML}\\

\bibliographystyle{mn2e}
\bibliography{ref}

\appendix

\section{Line Broadening from Beam Smearing} 
\label{app:line_broadening_from_beam_smearing}

Spectral line widths can be broadened wherever there is a large gradient in the rotation velocity on scales of the beam size.  This line broadening, commonly referred to as beam smearing, tends to have the largest effect near the centres of galaxies, where the rotation curve is steep, and can lead to significant increases in the line width of stacked profiles \citep[e.g.,][]{Stilp2013ApJ...765..136S,Ianjam2015AJ....150...47I,CalduPrimo2015AJ....149...76C}.

We require constraints on beam smearing when comparing line widths measured at different spatial resolutions in our data to distinguish whether broadened line profiles are the result of physical processes.  We estimate the maximum broadening from beam smearing by using a rolling tophat filter on the peak \hi velocity map to calculate the standard deviation over one beam.  This operation measures the beam-to-beam variation in the peak velocity field.  We note that these variations may not be entirely due to beam smearing and could arise from local variations in velocity, such as those measured for molecular cloud and envelope rotation in the \hi \citep{Imara:2011el}.  Therefore, our estimates are an {\it upper limit} on line broadening due to beam smearing.  This is a similar measurement to the approach used by \citet{CalduPrimo2016AJ....151...34C}, where they measure the width of the velocity distribution on local scales along the major and minor axes of M31.

\begin{figure}
\includegraphics[width=0.5\textwidth]{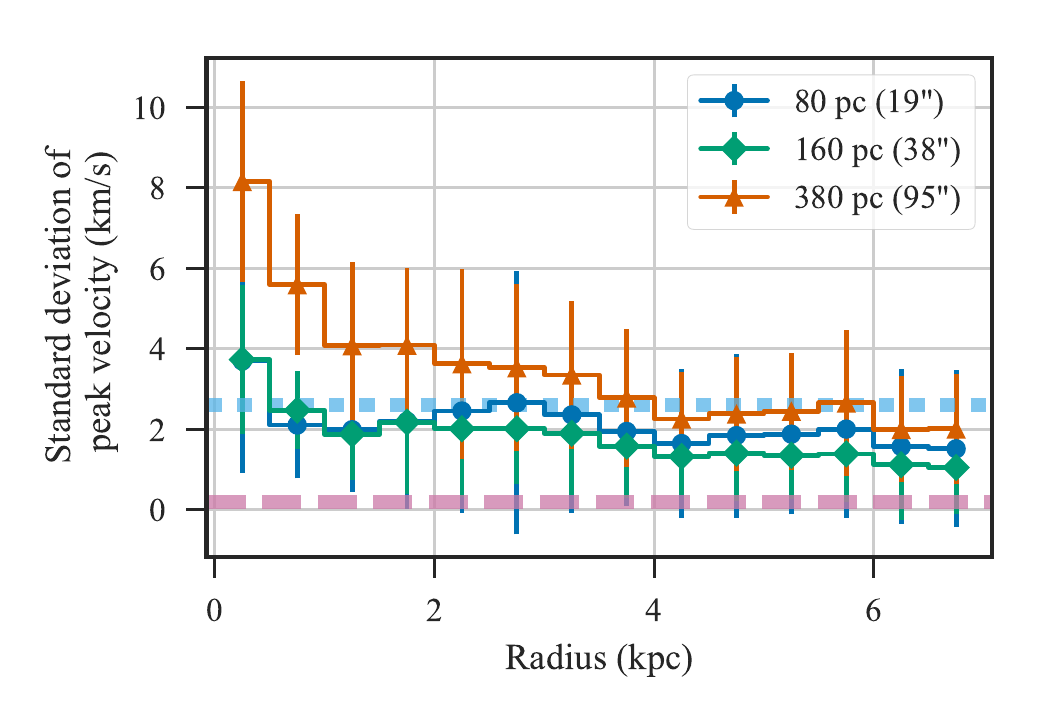}
\caption{\label{fig:beam_smear_radprofile} Average standard deviations from the peak \hi velocity map measured over one beam. Three average curves are shown measured within $0.5$ kpc bins at the original beam size (80~pc/$20\arcsec$; blue circles), and at twice (160~pc/$38\arcsec$; green diamonds) and five times (380~pc/$95\arcsec$; orange triangles) the original beam size. Error bars correspond to the standard deviation within each bin, uncorrected for the number of independent samples to demonstrate where the distributions are consistent with $0$ \kms.  The thick, horizontal lines correspond to the \hi (pink dashed; $0.2$~\kms) and \cotwoone (cyan dotted; $2.6$~\kms) channel widths.  The average values represent the maximum line broadening that could result from beam smearing.  Line widths at resolutions of 80 and 160~pc are uniformly broadened by $\sim2$~\kms, while the broadening at a resolution of 380~pc is $\sim3$~\kms and increases to $8$~\kms in the inner kpc.}
\end{figure}

We compute the standard deviation in the peak \hi velocity surface at the original (80~pc/$20\arcsec$) and degraded resolutions (160~pc/$38\arcsec$ and 380~pc/$95\arcsec$) used in Sections \ref{sub:total_profiles} \& \ref{sub:hi_line_widths_associated_with_co}.  Since large-scale variations in the peak \hi velocity describe the circular rotation curve, we calculate the average values of the standard deviation surfaces in $0.5$ kpc galactocentric radial bins.  If beam smearing significantly broadens the line, we expect the profile of average values to follow the derivative of the circular rotation curve, which is steepest within the inner 2 kpc of M33 \citepalias{Koch2018MNRAS}.  Figure \ref{fig:beam_smear_radprofile} shows the average of the standard deviation surfaces at the three different beam sizes used for the analysis.  The average radial profiles for beam widths of 80 and 160~pc do not have strong radial trends and show that beam smearing contributes at most $\sim2$~\kms to the line width.  We calculate the area-weighted average of the radial profiles in Figure \ref{fig:beam_smear_radprofile} and find values of $2.0^{+2.1}_{-1.8}$~\kms and $1.5^{+1.7}_{-0.8}$~\kms for beam sizes of 80 and 160~pc, respectively.  The uncertainties quoted here are the $15^{\rm th}$ and $85^{\rm th}$ percentiles of the radial profiles with the same area-weighted averaging applied.  Since the \cotwoone channel width is $2.6$~\kms, the line width broadening of \cotwoone from beam smearing is similar to the correction factor for the channel width.

Using the average line-of-sight \co line width of $4.3$~\kms at a resolution of 80~pc (Table \ref{tab:los_fitting}), the correction due to beam smearing gives a $\sim10$\% in the line width.

The average standard deviation profile measured at a beam size of 380~pc ($95\arcsec$) shows a strong radial trend within the inner 4~kpc, as expected from beam smearing.  The broadening from beam smearing is particularly strong within $R_{\rm gal} < 2$~kpc, where the maximum average standard deviation is $\sim8.2$ \kms.  The area-weighted average, as applied to the higher resolution measurements, is $2.8^{+1.0}_{-1.0}$~\kms.  Subtracting this mean value in quadrature from the 380~pc stacked line widths (Table \ref{tab:global_profiles}) gives corrected line widths of $8.4\pm0.9$~\kms for \hi and $6.7\pm3.6$~\kms for \cotwoone.  The \cotwoone is not constraining due to the uncertainty from the channel width, however the \hi line width range demonstrates that the $0.9$~\kms increase in the line width between the 160 and 380~pc data can entirely be explained by beam smearing.


\section{Forward-modelling the spectral response function}
\label{app:forward_model}

We forward-model the individual LOS fits to the \co data in \S\ref{sub:hi_line_widths_associated_with_co} with an approximation for the spectral response function.  Here, we briefly describe the fitting process.

We approximate the spectral response function in the \cotwoone IRAM-30m data based on the nearest neighbour channel correlation found by \citet[][$r=0.26$ for scales $>70$~pc, adjusted to a distance of 840~kpc used here]{Sun2018ApJ...860..172S}.  Using the empirically-derived relation from \citet{Leroy2016ApJ...831...16L}, this correlation corresponds to a channel-coupling factor of $k=0.11$ for a three-element Hann-like kernel ($[k, 1 - 2k, k]$), which we adopt as the spectral response function.

We forward-model the spectral response in two steps:

\begin{enumerate}
    \item The Gaussian model is sampled at the spectral channels of the observed spectra.  This sampling is equivalent to taking the weighted average of the Gaussian over the spectral channel width ($\Delta v$):
    \begin{equation}
    \label{eq:finite_gaussian}
    G(v) = \frac{A}{(\Delta v)^2} \left[ {\rm erf}\left( \frac{\mu - (v - \Delta v / 2)}{\sqrt{2}\sigma} \right) - {\rm erf}\left( \frac{\mu - (v + \Delta v / 2)}{\sqrt{2}\sigma} \right) \right],
    \end{equation}
    for a Gaussian centered at velocity $\mu$ with an amplitude of $A$ and width of $\sigma$ that is averaged over channels centered at $v$. The channel averaging is equivalent to convolving the Gaussian with a top-hat kernel with a width of $\Delta v$.

    \item The Gaussian sampled over the spectral channels is convolved with the Hann-like kernel described above.  This step accounts for the measured channel correlations in the observations.
\end{enumerate}

The sampled and convolved spectrum is then compared to the observed spectrum and the sum of the squared distances is the quantity minimized in the fit.  Using this approach removes biases in the fitted line width parameters \citep{Koch2018-linewidths}. This approach is similar to the forward-modelling in \citet{Rosolowsky2008ApJS..175..509R}.

We note that the non-linear least-squares fit used in this paper assumes that the data uncertainties are independent, which is not true due to the spectral-response of the data.  To test whether the parameter uncertainties from the covariance matrix of the fit are underestimated due to being correlated, we repeat the fitting procedure on 1000 simulated spectra sampled with channel widths of $\Delta v = \sigma$. White-noise is added to the spectra then convolved with the Hann-like kernel to give a peak S/N of 5.  We find that $\sim72\%$ of the fitted parameters are within the $1\mbox{--}\sigma$ uncertainty interval\footnote{An example of this test is available at \url{https://doi.org/10.5281/zenodo.1491796}}.  This is similar to the expected $68.2\%$ expected for a two-tailed p-test, demonstrating that the parameter uncertainties are not underestimated despite the correlated errors.

\section{Validating the Gaussian Decomposition} 
\label{app:validating_the_gaussian_decomposition}

We demonstrate our limited Gaussian decomposition method (\S\ref{sub:hi_line_widths_associated_with_co}) and perform two validation checks on the sample used in the analysis.

\begin{figure*}
\includegraphics[width=\textwidth]{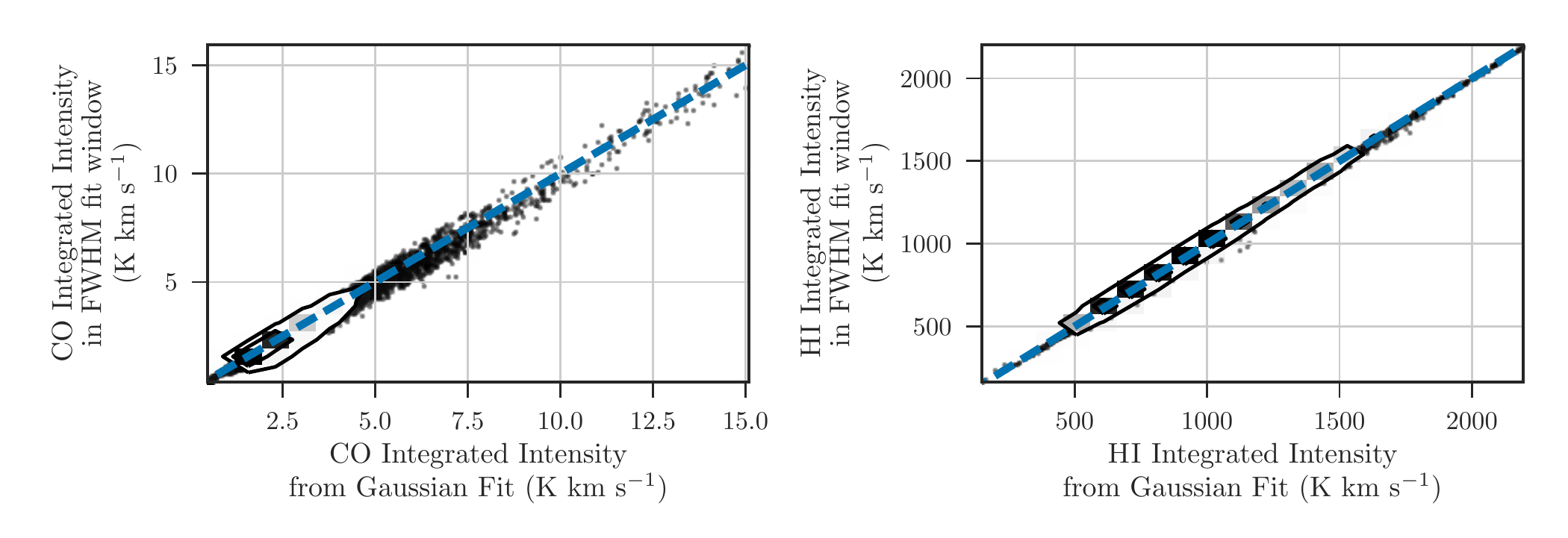}
\caption{\label{fig:coldens_fit_vs_fwhm_check} The \co (left) and \hi (right) integrated intensities from the spectra within the FWHM window used for the fit ($y$-axis) compared with the integral over the fitted Gaussian models ($x$-axis).  The blue-dashed line is the line of equality.  There is little deviation from the line of equality, indicating that the Gaussian models describe the data within the FWHM fitting windows well.}
\end{figure*}

The first check compares the surface densities from the integral over the fitted Gaussian model to the integrated intensity of the data located within the model's FWHM, scaled by $1 / {\rm erf}(\sqrt{2})$ to account for emission outside of the mask.  Figure~\ref{fig:coldens_fit_vs_fwhm_check} shows excellent agreement between the two methods for the \hi and \co\ fits.  This implies that the peaks are well-described by a Gaussian and validates the choice of model.

\begin{figure*}
\includegraphics[width=\textwidth]{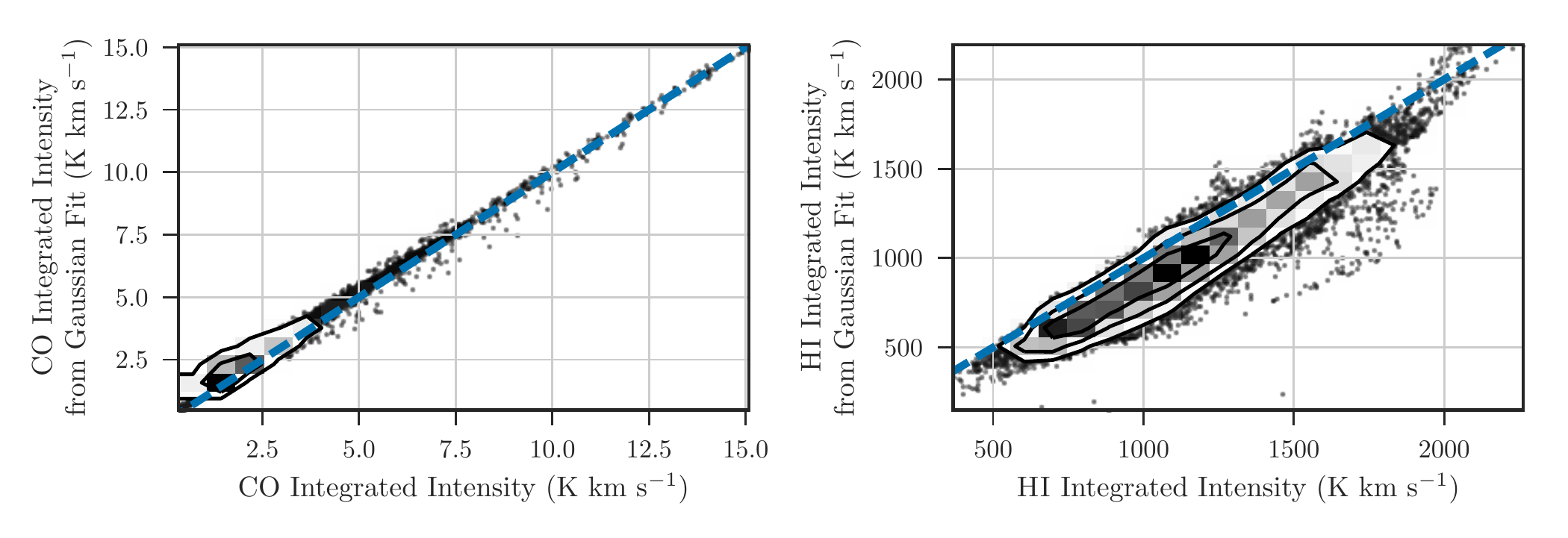}
\caption{\label{fig:coldens_fit_vs_moment_check} The \co (left) and \hi (right) integrated intensities from the integral over the fitted Gaussian models ($y$-axis) compared to the integrated intensity over the whole spectra ($x$-axis).  The blue-dashed line is the line of equality.  There remains good agreement for the \co spectra, however the \hi integrated intensity from the Gaussian fit is consistently smaller the from the entire spectra due to the additional \hi line structure.}
\end{figure*}

The second validation check is a comparison of the integral over the fitted Gaussian model with respect to the integrated intensities over the whole profile.  Figure~\ref{fig:coldens_fit_vs_moment_check} shows these quantities for the \co and \hi.  For the CO integrated intensities, there is no significant variation between the two quantities.  This is expected, since we require that the \co\ profiles be well-fit by a Gaussian in order to be in the sample.  The discrepancy between these quantities for the \hi is larger.  Again, this is expected, since the masking used in the \hi fitting is introduced to remove spectral features unlikely to be associated with the \co component.

With these two checks, we are confident that the clean sampled used for the analysis describes only single-peaked \co profiles and \hi profiles with a well-defined peak associated with the \co emission.

\subsection{Effect of \hi Masking on Fitting} 
\label{appsub:the_effect_of_hi_masking_on_fitting}

\begin{figure}
\includegraphics[width=0.5\textwidth]{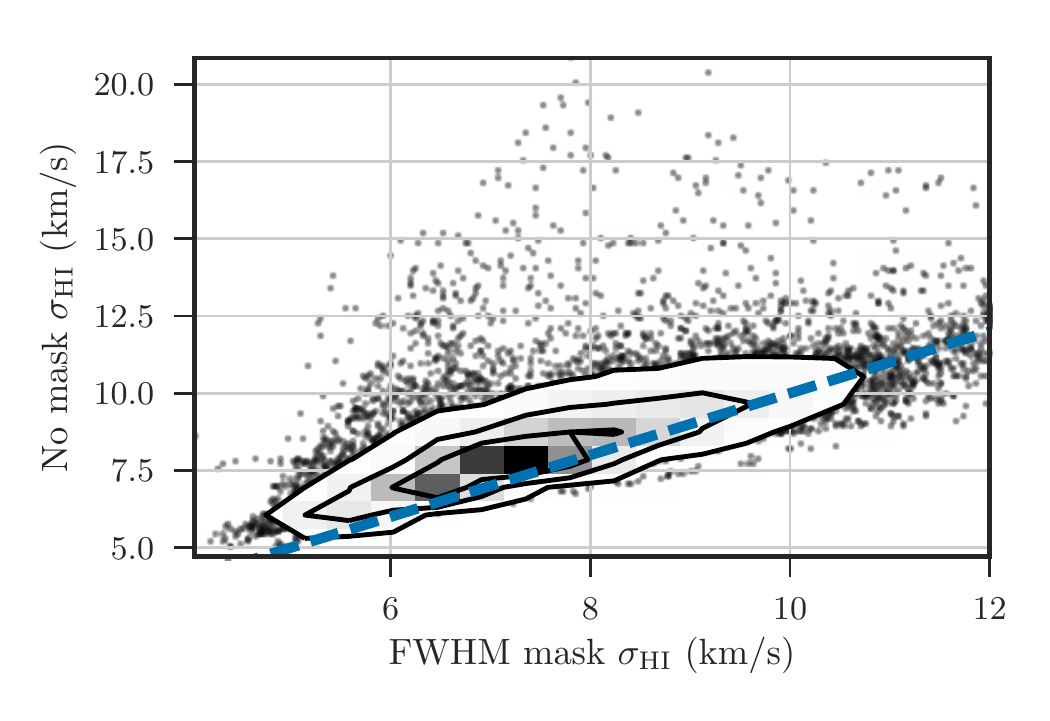}
\caption{\label{fig:sigma_w_wo_masking} \hi line widths fit with and without a FWHM mask around the peak. The contours represent the 2- to 4-$\sigma$ limits of the population, and points outside the contours are outliers beyond 4-$\sigma$.  The blue-dashed line indicates equality between the line widths.}
\end{figure}

We investigate how the FWHM mask affects the fitted \hi line width by repeating the fitting without the mask.  This procedure has been used in other studies relating \hi and \co line profiles from individual spectra \citep{Fukui2009ApJ...705..144F,Mogotsi2016AJ....151...15M}.  For \hi profiles with multiple components or prominent line wings, we expect that the fitted profiles without masking will be much wider. Figure~\ref{fig:sigma_w_wo_masking} shows that most \hi profiles are indeed wider without the masking, with the median width increasing from $7.4$ to $8.3$~\kms. This result highlights the need to carefully disinguish bright \hi emission from extended line wings to avoid biasing the \hi line widths and reinforces the use of the upper-limit of $12$ \kms in $\sigma_{\rm HI}$ set here to minimize contamination in our sample.

\subsection{Examples of Fitted Spectra} 
\label{appsub:examples_of_fitted_spectra}

\begin{figure}
\includegraphics[width=0.40\textwidth]{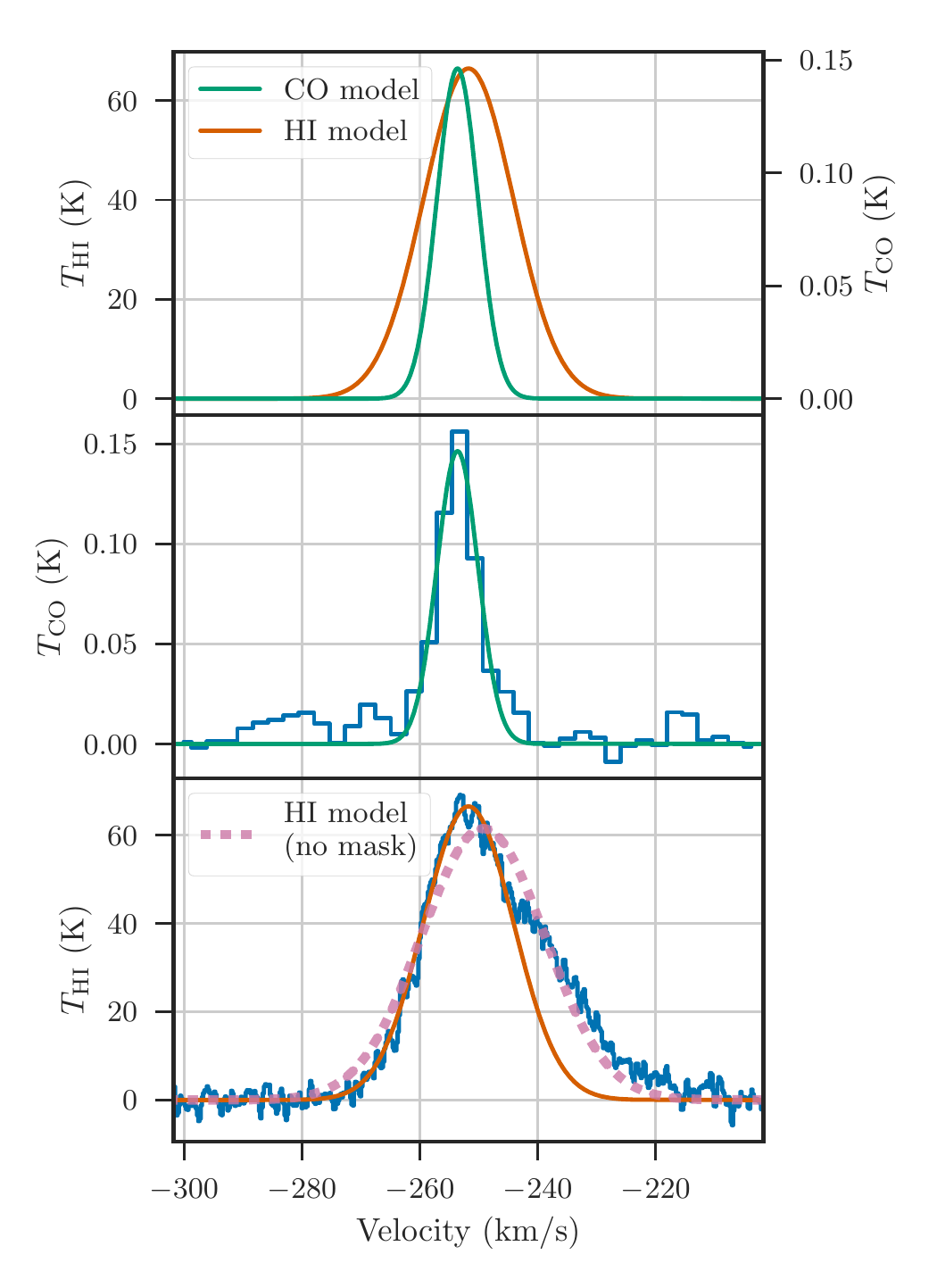}
\vspace{-3mm}
\caption{\label{fig:example_1} Example of the \hi and \co Gaussian fitting showing well-fit single components that are included in our clean sample.  The top panel shows the fitted \hi and \co profiles. The middle panel shows the \co spectrum with the fit overlaid. The bottom panel shows the same for the \hi data, and also includes a fit to the \hi data if no masking is applied when fitting (thick-dashed line; \S\ref{appsub:the_effect_of_hi_masking_on_fitting}). The \hi fit the brightest peak is significantly improved when masking is used.}
\end{figure}

\begin{figure}
\includegraphics[width=0.40\textwidth]{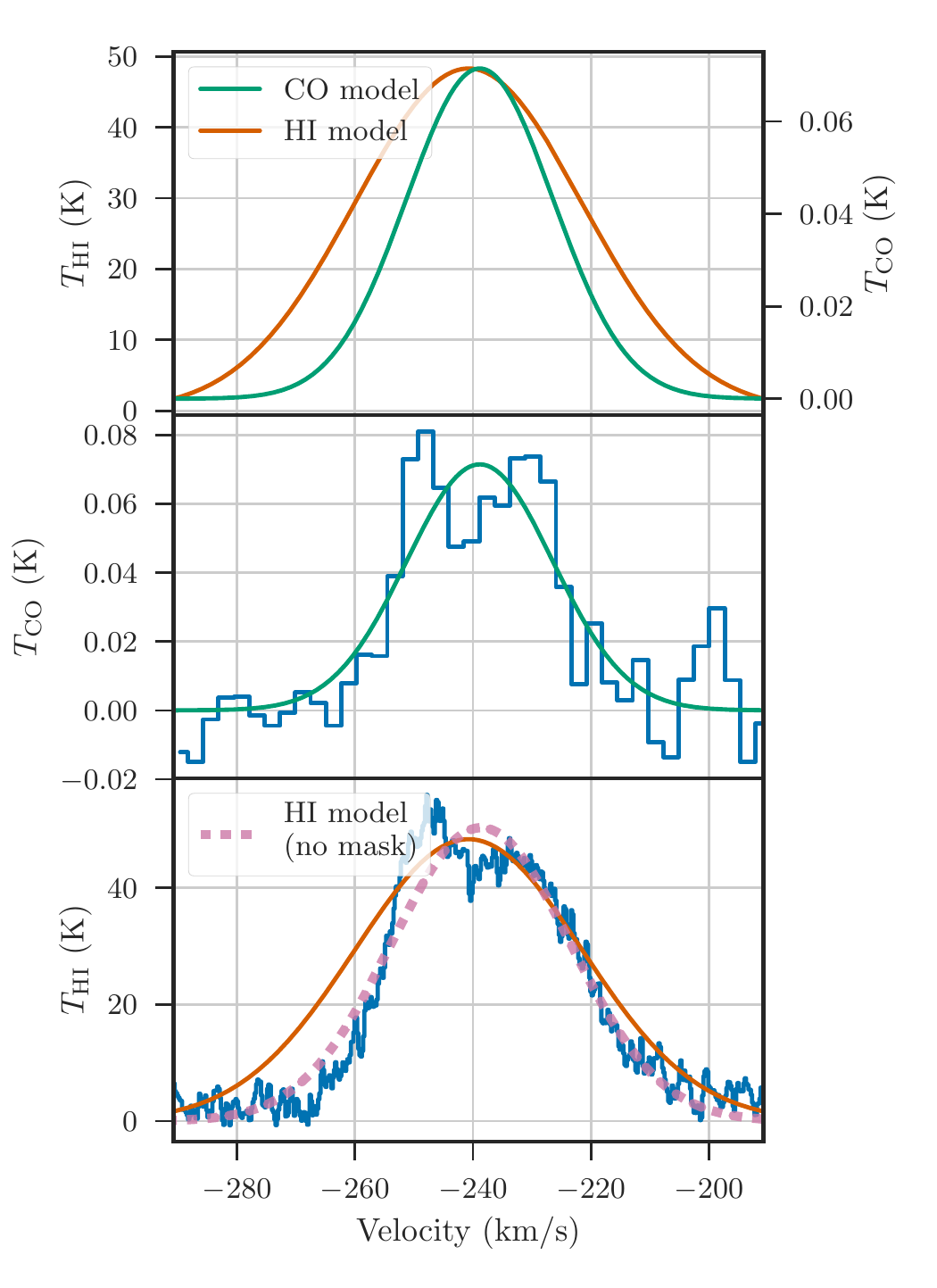}
\vspace{-3mm}
\caption{\label{fig:example_2} Same as Figure \ref{fig:example_1}.  This example shows failed fits in both tracers due to multiple-components and is rejected from the clean sample.  There appears to be two Gaussians in both spectra and fitting a multi-Gaussian model should distinguish between the two. Extending this analysis to multi-component spectra will be the focus of future work.}
\end{figure}

In Figures~\ref{fig:example_1} and \ref{fig:example_2}, we demonstrate fitted \hi and \co spectra and how our criteria for the analysis sample removes clear issues. Figure~\ref{fig:example_1} is an example of a valid fit to both \hi and \co, while Figure~\ref{fig:example_2} demonstrates a poor fit that is excluded from the analysis.

\section{Stacked Profile Widths} 
\label{app:stacked_profile_widths}

Table \ref{tab:hwhm_sigma_radial} shows the stacked line widths ($\sigma_{\rm HWHM}$) in radial bins.  These line widths are plotted in the top panel of Figure \ref{fig:radial_prof_widths} and described in \S\ref{subsub:radial_stacked_profiles}.

\begin{table}
\caption{\label{tab:hwhm_sigma_radial} HWHM line widths ($\sigma_{\rm HWHM}$) for the \hi and \cotwoone stacked profiles in 500~pc radial bins at 80~pc resolution. The uncertainties are propagated assuming an uncertainty of half the channel width and the uncertainty of each point in the spectrum is the standard deviation of values within that channel scaled by the square-root of the number of beams.}
\centering
\input{tables/hwhm_radial_widths.tex}
\end{table}


\label{lastpage}
\end{document}

%% file: tables/total_hwhm_fits_peakonly.tex
\begin{tabular}{lrr}
                     & \hi              & \cotwoone   \\\hline
\\[-1em]
\multicolumn{3}{c}{80 pc (20$''$) resolution}\\
\hline
$\sigma_{\rm HWHM}$ (\kms)            & $6.6\pm0.1$              & $4.6\pm0.9$     \\
\\[-1em]
$v_{\rm peak}$ (\kms)      & $0.0\pm0.1$              & $-0.2\pm0.9$     \\
\\[-1em]
$f_{\rm wings}$             & $0.25_{-0.01}^{+0.01}$     & $0.21_{-0.10}^{+0.12}$ \\
\\[-1em]
$\sigma_{\rm wings}$ (\kms) & $24.0_{-0.4}^{+0.3}$       & $18_{-3}^{+5}$ \\
\\[-1em]
$a$                        & $0.021_{-0.005}^{+0.014}$  & $-0.05_{-0.14}^{+0.09}$ \\
\\[-1em]
$\kappa$                   & $-0.059_{-0.003}^{+0.004}$ & $0.02_{-0.09}^{+0.06}$ \\
\hline
\multicolumn{3}{c}{160 pc (38$''$) resolution}\\
\hline
$\sigma_{\rm HWHM}$ (\kms) & $8.0\pm0.1$              & $5.9\pm0.9$     \\
\\[-1em]
$v_{\rm peak}$ (\kms)      & $-0.1\pm0.1$              & $-0.4\pm0.9$     \\
\\[-1em]
$f_{\rm wings}$             & $0.19_{-0.01}^{+0.01}$     & $0.14_{-0.09}^{+0.10}$ \\
\\[-1em]
$\sigma_{\rm wings}$ (\kms) & $29.3_{-0.4}^{+0.3}$       & $20_{-4}^{+8}$ \\
\\[-1em]
$a$                        & $0.012_{-0.012}^{+0.007}$  & $-0.07_{-0.10}^{+0.20}$ \\
\\[-1em]
$\kappa$                   & $-0.022_{-0.003}^{+0.004}$ & $0.00_{-0.06}^{+0.05}$ \\
\hline
\multicolumn{3}{c}{380 pc (95$''$) resolution}\\
\hline
$\sigma_{\rm HWHM}$ (\kms) & $8.9\pm0.1$              & $7.2\pm0.9$     \\
\\[-1em]
$v_{\rm peak}$ (\kms)      & $0.0\pm0.1$              & $-0.2\pm0.9$     \\
\\[-1em]
$f_{\rm wings}$             & $0.19_{-0.01}^{+0.01}$     & $0.15_{-0.07}^{+0.09}$ \\
\\[-1em]
$\sigma_{\rm wings}$ (\kms) & $32.1_{-0.4}^{+0.3}$       & $22_{-3}^{+4}$ \\
\\[-1em]
$a$                        & $0.033_{-0.015}^{+0.010}$  & $-0.07_{-0.10}^{+0.20}$ \\
\\[-1em]
$\kappa$                   & $-0.035_{-0.035}^{+0.004}$ & $-0.01_{-0.05}^{+0.05}$ \\
\end{tabular}

%% file: tables/los_scale_linewidth.tex
\begin{tabular}{crrr}
Resolution (pc) & \multicolumn{2}{c}{$\sigma$ (\kms)} & Fitted $\sigma_{\rm CO} / \sigma_{\rm HI}$\\\hline
\\[-1em]
                     & \hi              & \cotwoone &   \\\hline
\\[-1em]
80 (20$''$)    & $7.4^{+1.7}_{-1.3}$ & $4.3^{+1.5}_{-1.0}$ & $0.56\pm0.01$    \\
\\[-1em]
160 (38$''$) & $8.4^{+1.8}_{-1.2}$ & $5.0^{+1.4}_{-1.1}$ & $0.57\pm0.01$ \\
\\[-1em]
380 (95$''$) & $11.0^{+2.7}_{-2.0}$ & $7.3^{+2.4}_{-1.6}$ & $0.63\pm0.01$ \\
\\[-1em]
\end{tabular}

%% file: tables/total_hwhm_los_stacking.tex
\begin{tabular}{lrr}
                     & \hi              & \cotwoone   \\\hline
\\[-1em]
\multicolumn{3}{c}{(i)\ Fitted model components aligned to CO Model $v_{0}$}\\
\hline
$\sigma_{\rm HWHM}$ (\kms)  & $7.6\pm0.1$              & $4.2\pm0.9$     \\
\\[-1em]
$f_{\rm wings}$             & $0.03_{-0.01}^{+0.01}$     & $0.05_{-0.13}^{+0.17}$ \\
\\[-1em]
\hline
\multicolumn{3}{c}{(ii)\ Fitted model components aligned to \hi Model $v_{0}$}\\
\hline
$\sigma_{\rm HWHM}$ (\kms)  & $7.4\pm0.1$              & $4.6\pm0.9$     \\
\\[-1em]
$f_{\rm wings}$             & $0.03_{-0.01}^{+0.01}$     & $0.03_{-0.12}^{+0.16}$ \\
\\[-1em]
\hline
\multicolumn{3}{c}{(iii)\ LOS spectra aligned to \hi Model $v_{0}$}\\
\hline
$\sigma_{\rm HWHM}$ (\kms) & $7.4\pm0.1$              & $4.8\pm0.9$     \\
\\[-1em]
$f_{\rm wings}$             & $0.19_{-0.01}^{+0.01}$     & $0.07_{-0.12}^{+0.10}$ \\
\\[-1em]
\hline
\multicolumn{3}{c}{(iv)\ LOS Spectra aligned to \hi $v_{\rm peak}$}\\
\hline
$\sigma_{\rm HWHM}$ (\kms) & $7.3\pm0.1$              & $4.7\pm0.9$     \\
\\[-1em]
$f_{\rm wings}$             & $0.20_{-0.01}^{+0.01}$     & $0.08_{-0.12}^{+0.15}$ \\
\\[-1em]
\hline
\multicolumn{3}{c}{(v)\ All LOS spectra aligned to \hi $v_{\rm peak}$}\\
\hline
$\sigma_{\rm HWHM}$ (\kms) & $7.6\pm0.1$              & $4.8\pm0.9$     \\
\\[-1em]
$f_{\rm wings}$             & $0.22_{-0.01}^{+0.01}$     & $0.12_{-0.11}^{+0.14}$ \\
\\[-1em]
\end{tabular}

%% file: tables/hwhm_radial_widths.tex
\begin{tabular}{lcc}
$R_{\rm gal}$ (kpc)         & $\sigma_{\rm HI}$ (\kms)    & $\sigma_{\rm CO}$ (\kms)   \\\hline
\\[-1em]
$0.0\mbox{--}0.5$           & $8.0\pm0.1$           & $5.0\pm0.9$         \\
$0.5\mbox{--}1.0$           & $7.3\pm0.1$           & $4.6\pm1.0$         \\
$1.0\mbox{--}1.5$           & $6.9\pm0.1$           & $4.5\pm1.0$         \\
$1.5\mbox{--}2.0$           & $7.3\pm0.1$           & $4.8\pm0.9$         \\
$2.0\mbox{--}2.5$           & $7.4\pm0.1$           & $4.8\pm0.9$         \\
$2.5\mbox{--}3.0$           & $7.4\pm0.1$           & $4.7\pm0.9$         \\
$3.0\mbox{--}3.5$           & $6.8\pm0.1$           & $4.7\pm0.9$         \\
$3.5\mbox{--}4.0$           & $7.2\pm0.1$           & $4.8\pm1.0$         \\
$4.0\mbox{--}4.5$           & $6.7\pm0.1$           & $4.4\pm1.0$         \\
$4.5\mbox{--}5.0$           & $6.9\pm0.1$           & $4.6\pm1.0$         \\
$5.0\mbox{--}5.5$           & $6.9\pm0.1$           & $4.1\pm1.0$         \\
$5.5\mbox{--}6.0$           & $6.9\pm0.1$           & $3.9\pm0.9$         \\
$6.0\mbox{--}6.5$           & $6.6\pm0.1$           & $4.3\pm1.0$         \\
$6.5\mbox{--}7.0$           & $6.1\pm0.1$           & $3.8\pm1.1$         \\
\end{tabular}